\def\year{2022}\relax
\documentclass[letterpaper]{article} % DO NOT CHANGE THIS
\usepackage{aaai22}  % DO NOT CHANGE THIS
\usepackage{times}  % DO NOT CHANGE THIS
\usepackage{helvet}  % DO NOT CHANGE THIS
\usepackage{courier}  % DO NOT CHANGE THIS
\usepackage[hyphens]{url}  % DO NOT CHANGE THIS
\usepackage{graphicx} % DO NOT CHANGE THIS
\usepackage{natbib}  % DO NOT CHANGE THIS AND DO NOT ADD ANY OPTIONS TO IT
\usepackage{caption} % DO NOT CHANGE THIS AND DO NOT ADD ANY OPTIONS TO IT
\DeclareCaptionStyle{ruled}{labelfont=normalfont,labelsep=colon,strut=off} % DO NOT CHANGE THIS
\usepackage{algorithm}
\usepackage{algorithmic}

\usepackage{newfloat}
\usepackage{listings}
\floatstyle{ruled}
\newfloat{listing}{tb}{lst}{}
\floatname{listing}{Listing}
\usepackage{hyperref}       % hyperlinks
\usepackage{url}            % simple URL typesetting
\usepackage{booktabs}       % professional-quality tables
\usepackage{amsfonts}       % blackboard math symbols
\usepackage{nicefrac}       % compact symbols for 1/2, etc.
\usepackage{microtype}      % microtypography
\usepackage{xcolor}         % colors

\usepackage[utf8]{inputenc} % allow utf-8 input
\usepackage[T1]{fontenc}    % use 8-bit T1 fonts
\usepackage{booktabs}       % professional-quality tables
\usepackage{amsfonts}       % blackboard math symbols
\usepackage{nicefrac}       % compact symbols for 1/2, etc.
\usepackage{microtype}      % microtypography

\usepackage{amsmath}
\usepackage{amsthm}
\usepackage{xcolor}
\usepackage{algorithm}
\usepackage{algorithmic}

\usepackage{graphicx}
\usepackage{subcaption}
\usepackage[export]{adjustbox}
\usepackage{enumitem}
\usepackage{wrapfig}

\usepackage[title]{appendix}
\newtheorem{theorem}{Theorem}

\newtheorem{definition}{Definition}
\newtheorem{lemma}{Lemma}
\newtheorem{corollary}{Corollary}
\newtheorem{proposition}{Proposition}

\title{Empirical Policy Optimization for $n$-Player Markov Games}
\author{
    Yuanheng~Zhu\textsuperscript{\rm 1}, Dongbin~Zhao\textsuperscript{\rm 1}, Mengchen~Zhao\textsuperscript{\rm 2}, Dong~Li\textsuperscript{\rm 2}
}
\affiliations{
    \textsuperscript{\rm 1} State Key Laboratory of Management and Control for Complex Systems, Institute of Automation, Chinese Academy of Sciences, Beijing 100190, China \\
    \textsuperscript{\rm 2} Huawei Noah's Ark Lab, Beijing, China \\
    yuanheng.zhu@ia.ac.cn, dongbin.zhao@ia.ac.cn, zhaomengchen@huawei.com, lidong106@huawei.com
}

\usepackage{bibentry}
\begin{document}

\maketitle

\begin{abstract}
In single-agent Markov decision processes, an agent can optimize its policy based on the interaction with environment. In multi-player Markov games (MGs), however, the interaction is non-stationary due to the behaviors of other players, so the agent has no fixed optimization objective. In this paper, we treat the evolution of player policies as a dynamical process and propose a novel learning scheme for Nash equilibrium.
The core is to evolve one's policy according to not just its current in-game performance, but an aggregation of its performance over history. We show that for a variety of MGs, players in our learning scheme will provably converge to a point that is an approximation to Nash equilibrium. Combined with neural networks, we develop the \emph{empirical policy optimization} algorithm, that is implemented in a reinforcement-learning framework and runs in a distributed way, with each player optimizing its policy based on own observations. We use two numerical examples to validate the convergence property on small-scale MGs with $n\ge 2$ players, and a pong example to show the potential of our algorithm on large games.
\end{abstract}

\section{Introduction}

Markov games (MGs), or stochastic games termed in some references \citep{Shapley1095, hu2003nash}, are the extension of Markov decision processes (MDPs) from single-agent environment to multi-player scenarios \citep{littman1994markov, julien2015app}.
Compared to normal-form games (NFGs) \citep{acar2020dis, li2021sub} that are stateless and lack the transition of states, MG players encounter multiple decision moments in one round, and at each step have to take into account current game states to make decisions. Each player aims to maximize its sum of rewards over time horizon, instead of one-stage payoff in NFGs.
Another famous type of sequential games is extensive-form games (EFGs) \citep{heinrich2015fictitious, zhou2020posterior, brown2020com}, in which the moves of different players are played in orders, compared to MG players acting simultaneously. EFGs use a game tree to describe the game process, and reshaping MGs to EFGs results in exponential blowup in size with respect to horizon.

In game theory, an important concept is Nash equilibrium, at which no player has the intention of deviating its strategy without sacrificing current payoff. However, even for simple NFGs, computing Nash equilibrium is proved to be PPAD-complete \citep{das2009the, liu2021on}. Alternatively, learning-based schemes provide a computational-intelligence way to approach equilibriums and have now become appealing to researchers \citep{leslie2005ind, coucheney2015penalty, gao2021on}.

Deep reinforcement learning (DRL) is a powerful tool in sequential decision makings \citep{mnih2015human, Schulman2017ProximalPO}, and has also received attention from the game field. One biggest challenge of DRL in MGs is the non-stationarity of optimization objectives, since each player's payoff is determined by others' behaviours. Recent progress has been made on two-player zero-sum cases \citep{Silver2017Mastering, lockhart2019comp, dask2020ind}. For more general $n$-player games with arbitrary number of players, one solution is to convert MGs to empirical games (seen as NFGs) with every policy being a strategy option \citep{lanctot2017a}. Then, DRL optimizes the policy of each player in a game environment against a group of fixed opponents, whose strategies are mixtures of empirical policies \citep{david2019open, mcaleer2020pipeline, Muller2020A}. However, this approach is computationally expensive because different players are trained in different game environments with specific opponents, and the synthesis of opponent strategies may require additional computational efforts.

In recent years, reinforcement learning (RL) has promoted the development of payoff-based learning scheme in NFGs \citep{coucheney2015penalty, perkins2017mixed, mer2019learning, gao2021on}. Neglecting what the others' strategies are, each player updates its strategy based on the aggregation of its on-going payoffs. If all players follow the same update rule, the evolution of their strategies converges to a Nash distribution with theoretical guarantees. However, in the field of MGs, this scheme faces obstacles because the strategy becomes a policy mapping from states to action distributions and the payoff is the expectation of sum of future rewards. Motivated by that, in this paper we establish a \emph{continuous-time learning dynamics} (CTLD) for arbitrary $n$-player MGs. Instead of changing the form of games, all players in CTLD interact in the same game environment and each player evolves its policy based on the aggregation of its on-going performance. To facilitate large-scale applications, an \emph{empirical policy optimization} (EPO) algorithm is developed. Player policies are represented by neural networks (NNs) and the parameters are trained based on the whole history of experience in an RL way. Compared to existing methods, our contributions are three-fold:
\begin{itemize}[itemsep=0pt,topsep =0pt,leftmargin=*]
  \item The learning scheme runs in a totally distributed way and players require no other game information but only own observations. Players do not need to know how many players are participating and what strategies the others are playing, so the scheme is applicable to arbitrary $n$-player cases.
  \item Based on the fixed-point theorem and Lyapunov stability of dynamical systems, we prove that CTLD converges to a Nash distribution (an approximation to Nash equilibrium with arbitrary precision) for a variety of MGs.
  \item The simplicity and distributed property makes the learning scheme compatible with RL framework for large-scale games. Experimental results show the efficiency of EPO in approaching Nash equilibrium.
\end{itemize}

\section{Related work}

Early research on MGs was mainly value-based methods that aimed to solve Nash values of Bellman-like equations \citep{littman1994markov}. If the game model is known, one can apply dynamic programming (e.g. \citep{lag2002learning, lag2002value}). Otherwise, one can learn the values based on online observations like Q-learning \citep{hu2003nash, zhu2020online}. There is also recent progress on sample complexity \citep{wei2017online, zhang2020model, bai2020near} and risk awareness \citep{huang2020model, tian2021bounded}. However, value-based methods rely on the Nash computing (of NFGs) at every state, and the optimization over joint-action space suffers from combinatorial explosion as the number of players increases.

Policy optimization, or policy update is efficient in optimizing agent policies \citep{schulman2015trust, Schulman2017ProximalPO}. When applying to multi-player scenarios, extra efforts are needed to manipulate the update direction towards Nash equilibrium. As a special case of MGs, two-player zero-sum games have received much attention. \citet{srin2018actor} showed that when directly applying independent policy update rules in zero-sum sequential games, the regret had no sublinearity in iterations, in other words the process may not converge to a Nash equilibrium. \citet{lockhart2019comp} improved the results by optimizing one player's policy against its best-response opponent, and proved when using counterfactual values, the joint policies converged to a Nash equilibrium in two-player EFGs. A similar concept was adopted by \citet{zhang2019policy} for zero-sum linear quadratic games, and achieved the global convergence results. \citet{dask2020ind} chose a two-timescale learning rates for the independent learning of min-player and max-player, which can be seen as a softened ``gradient descent vs. best response'' scheme.

For more general $n$-player games, the development is limited. A recent progress is policy-search response oracles (PSRO), which was first proposed by \citet{lanctot2017a}. The main idea is to reduce MGs to empirical games, or meta-games, whose policy sets are composed of empirical policies in history. Each player finds a (approximate) best response to its opponents' meta-strategies, and the new policies are added into policy sets for the next iteration. The advantage of PSRO is that it provides a unified framework for different choices of meta-solvers. \citet{david2019open} proposed to use rectified Nash mixtures to encourage policy diversity. \citet{Muller2020A} introduced the $\alpha$-rank multi-agent evaluation metric \citep{omidshafiei2019alpha} in PSRO, and showed promising performance in computing equilibria. However, PSRO is intensive in computation from two aspects: 1) the policy update of each player is separated in different game environments with different opponents; 2) additional computational efforts are required by meta-solvers and empirical payoff evaluation.

\section{Background and terminology}

A Markov game played by a finite set of players $\mathcal{N} = \{1, 2, \dots, n\}$ can be described by $\mathcal{MG} = ( \mathcal{N}, \mathcal{S}, \{\mathcal{A}^i \}_{i \in \mathcal{N}}, \{\mathcal{R}^i \}_{i \in \mathcal{N}}, \mathcal{P}, \gamma, \rho_0 )$, where $\mathcal{S}$ is the finite set of states; $\mathcal{A}^i$ is the finite set of actions for each player $i \in \mathcal{N}$; $\mathcal{R}^i: \mathcal{S} \times \{\mathcal{A}^i \} \to \mathbb{R}$ is player $i$'s (bounded) reward function; $\mathcal{P}: \mathcal{S} \times \{\mathcal{A}^i \} \times \mathcal{S} \to [0,1]$ is the state transition function; $\gamma \in (0,1)$ is the discounted factor; $\rho_0$ is the initial state distribution.

%At the beginning of each round, the game initializes a state $s_0 \sim \rho_0$ and repeats the following four steps until terminal: (i) all players observe the current state $s_k \in \mathcal{S}$; (ii) each player $i$ independently chooses an action $a^i_k \in \mathcal{A}^i$ and simultaneously executes in the game; (iii) each player realizes its reward signal $r^i_k \sim \mathcal{R}^i (s_k, \boldsymbol{a}_k)$ according to the current $s_k$ and the joint action $\boldsymbol{a}_k = (a^i_k)_{i \in \mathcal{N}}$; and (iv) the game transitions to the next state by $s_{k+1} \sim \mathcal{P}(s_k, \boldsymbol{a}_k)$.

In the field of RL, one is interested in a \emph{policy} $\pi\colon \mathcal{S} \times \mathcal{A} \to [0,1]$ that describes the action selection probability at given $s$ by $\pi(\cdot | s) \in \Delta (|\mathcal{A}|)$. $\Delta (|\mathcal{A}|)$ denotes the simplex $ \{ p \in \mathbb{R}^{|\mathcal{A}|} \mid p \ge 0 \text{ componentwise, and }  \boldsymbol{1}^T p =1  \}$. Assuming each player has an independent $\pi^i \colon \mathcal{S} \times \mathcal{A}^i \to [0,1]$, the aggregation forms the policy profile $\boldsymbol{\pi} = (\pi^i)_{i \in \mathcal{N}}$, and player $i$'s expected return, or \emph{value}, starting from $s_0$ is defined as
$
  V_{\boldsymbol{\pi}}^i (s_0) = \mathbb{E} \big[ \sum_{k=0}^{\infty} \gamma^k r^i_{k} \mid \boldsymbol{a}_k \sim \boldsymbol{\pi} (s_k), r^i_k \sim \mathcal{R}^i(s_k, \boldsymbol{a}_k), s_{k} \sim \mathcal{P}(s_k, \boldsymbol{a}_k) \big]
$. Another important RL concept is state-action value, or \emph{Q} value:
$
  Q_{\boldsymbol{\pi}}^i (s, a^i) = \mathbb{E} \big[ r^i + \gamma V_{\boldsymbol{\pi}}^i (s') \mid \boldsymbol{a}^{-i} \sim \boldsymbol{\pi}^{-i}(s), r^i \sim \mathcal{R}^i(s, \boldsymbol{a}^{i}), s' \sim \mathcal{P}(s, \boldsymbol{a}) \big]
$. We use $-i$ to indicate the other players in $\mathcal{N}$ except $i$. The difference between $Q_{\boldsymbol{\pi}}^i$ and $V_{\boldsymbol{\pi}}^i$ is known as the \emph{advantage}:
$
  A^i_{\boldsymbol{\pi}} (s,a^i) = Q_{\boldsymbol{\pi}}^i (s, a^i) - V_{\boldsymbol{\pi}}^i (s)
$. In what follows, we sometimes use $A_{\pi^i, \boldsymbol{\pi}^{-i}}$ to denote the observed advantage of player $i$ when it is playing $\pi^i$ and the others are playing $\boldsymbol{\pi}^{-i}$.

Given the initial state distribution $\rho_0$, player $i$'s \emph{payoff} is the expected value under the profile $\boldsymbol{\pi}$:
$
  u^i(\pi^i , \boldsymbol{\pi}^{-i}) = \sum_{s} \rho_0(s) V^i_{\boldsymbol{\pi}} (s)
$, and each player aims to maximize its own payoff. Once the other policies $\boldsymbol{\pi}^{-i}$ are fixed, the game is reduced to player $i$'s MDP, and the difference in performance between player $i$'s any two policies $\pi^i$ and $\pi^i_{\dagger}$ follows the policy update in \citep{kakade2002appr, schulman2015trust}. Before restating the lemma in game setting, we let $
  \rho_{\boldsymbol{\pi}} (s) = \left( P(s_0 = s) + \gamma P(s_1 = s)+ \gamma^2 P(s_2=s) + \dots \right)
$ be the \emph{discounted visitation frequencies}, where $s_0 \sim \rho_0$ and all players follow $\boldsymbol{\pi}$.

\begin{lemma}[Restatement of Policy Update \citep{kakade2002appr, schulman2015trust}]
\label{lem:policy_update}
  Given the other policy profile $\boldsymbol{\pi}^{-i}$, player $i$'s payoffs under two policies $\pi^i$ and $\pi^i_{\dagger}$ satisfy
  \begin{multline*}
    u^i (\pi^i_{\dagger}, \boldsymbol{\pi}^{-i}) = u^i (\pi^i, \boldsymbol{\pi}^{-i}) \\
     + \sum_{s} \rho_{\pi^i_{\dagger}, \boldsymbol{\pi}^{-i}} (s) \sum_{a^i} \pi^i_{\dagger} (s,a^i) A_{\pi^i, \boldsymbol{\pi}^{-i}} (s,a^i).
  \end{multline*}
\end{lemma}

Player $i$'s \emph{best response} to a given profile $\boldsymbol{\pi}^{-i}$ is the policy that maximizes its payoff:
$
  \beta^i (\boldsymbol{\pi}^{-i}) = \arg\max_{\pi^i \in \Pi^i} u^i(\pi^i , \boldsymbol{\pi}^{-i})
$, where $\Pi^i$ represents the policy space of player $i$. If in a profile $\boldsymbol{\pi}_* = ( {\pi}^{i}_*)_{i \in \mathcal{N}}$ each policy is the best response of the others, the profile is called the \emph{Nash equilibrium} and satisfies
$
  u^i({\pi}^{i}_* , \boldsymbol{\pi}^{-i}_*) \ge u^i({\pi}^{i} , \boldsymbol{\pi}^{-i}_*)$, $\forall \pi^i \in \Pi^i$ and $\forall i \in \mathcal{N}$.
%  The following lemma describes the relationship between the Nash equilibrium policy and any policies in terms of the sum of per-timestep advantages. The proof is presented in Appendices to save space.
%\begin{lemma}
%  \label{lem:nash_equil_ineq}
%  If $\boldsymbol{\pi}_* $ is a Nash equilibrium, then for any $\pi^i \in \Pi^i$, $\forall i \in \mathcal{N}$,
%  \begin{multline*}
%    \sum_s \rho_{\boldsymbol{\pi}_*} (s) \sum_{a^i} \pi^{i}_* (s,a^i) A^i_{\boldsymbol{\pi}_*} (s,a^i) \ge \\
%    \sum_s \rho_{\boldsymbol{\pi}_*} (s) \sum_{a^i} \pi^i (s,a^i) A^i_{\boldsymbol{\pi}_*} (s,a^i) .
%  \end{multline*}
%\end{lemma}

For any profile in the joint policy space $\boldsymbol{\Pi} = (\times \Pi^i)_{i \in \mathcal{N}}$, $\mathrm{NashConv}$ provides a metric to measure the distance to Nash, i.e. $\mathrm{NashConv} (\boldsymbol{\pi}) = \sum_i \max_{\pi} u^i(\pi, \boldsymbol{\pi^{-i}}) - u^i(\pi^i, \boldsymbol{\pi^{-i}}) $. It always has $\mathrm{NashConv} (\boldsymbol{\pi}) \ge 0$ for any profile and is equal to zero at the Nash equilibrium.

\begin{figure}
\centering
\includegraphics[width=0.25\textwidth]{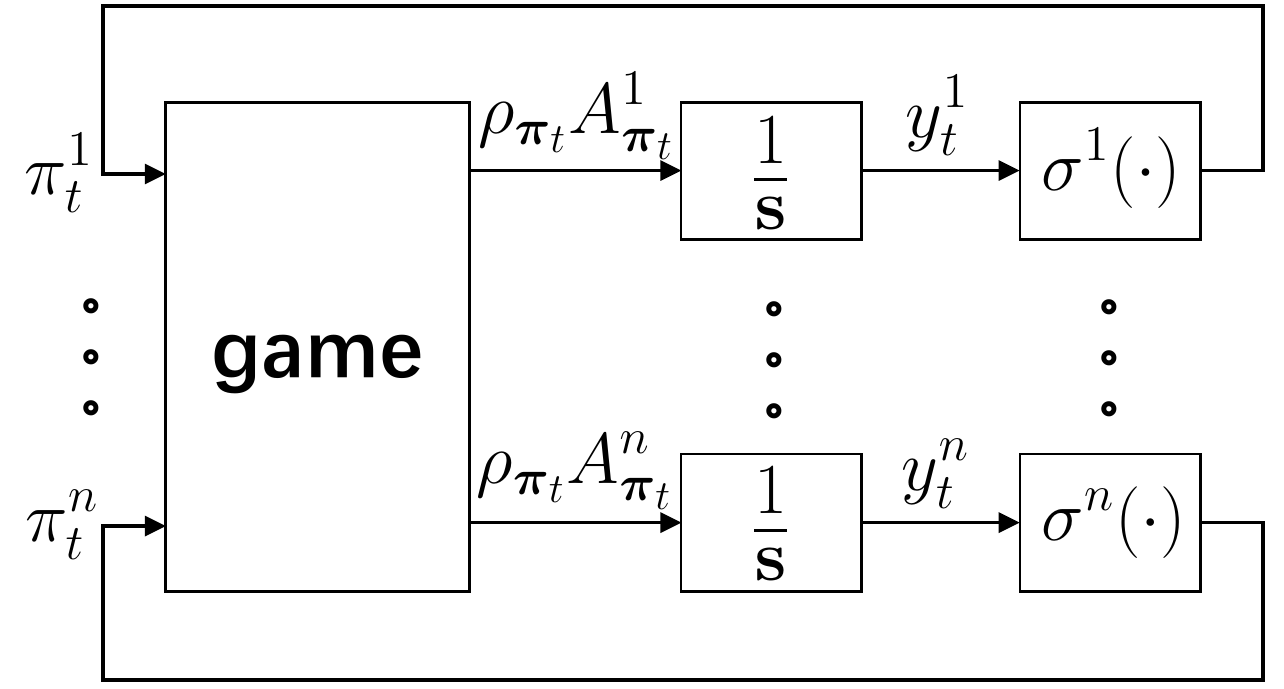}
\caption{Block diagram of CTLD. $\frac{1}{\mathbf{s}}$ indicates the integrator block. }
\label{fig_block_diagram}
\end{figure}

\section{Convergent continuous-time learning scheme}

\subsection{Continuous-time learning dynamics}

We now establish the learning scheme for the Nash equilibrium of $n$-player MGs. The outline is that each player keeps a score function that records its on-going performance, and then maps the score to a policy that is played with the others to evaluate the performance. The process is modeled in continuous time, repeated with an infinitesimal time step between three stages described below. A block diagram of the dynamical system is given in Figure \ref{fig_block_diagram}.

\textbf{\emph{1) Assessment Stage:}} Consider the current time $t$ and all players' profile $\boldsymbol{\pi}_t = (\pi^i_t)_{i \in \mathcal{N}}$. Player $i$'s \emph{score} $y^i_t$ keeps the running average of past weighted advantages $\rho_{\boldsymbol{\pi}_{\tau}} (s)  A^i_{\boldsymbol{\pi}_{\tau}} (s,a^i)$, $\tau \in [0, t)$, at every state-action pair, based on the exponential discounting aggregation
\begin{multline*}
  y^i_t(s,a^i) = e^{-\eta t} y^i_0(s,a^i) + \\
  \eta \int_0^t e^{-\eta (t-\tau)} \rho_{\boldsymbol{\pi}_{\tau}} (s)  A^i_{\boldsymbol{\pi}_{\tau}} (s,a^i) d\tau, \quad \forall s \in \mathcal{S}, \forall a^i \in \mathcal{A}^i
\end{multline*}
where $\eta > 0$ is the learning rate and $y^i_0$ is an arbitrary starting point. By formally defining an operator $w^i$ for each player mapping from policy to weighted advantage: $
  [w^i (\boldsymbol{\pi})] (s,a^i) = \rho_{\boldsymbol{\pi}} (s) A^i_{\boldsymbol{\pi}} (s,a^i)
$, the evolution of score can be described in differential form\footnote{One should distinguish between the two time indices $t$ and $k$: the former indicates the evolution of score or policy at higher function level, while the latter indicates the game transition at lower state level.}
\begin{equation}
\label{eq_yi_dynamics}
  \dot{y}^i_t  = \eta \left( w^i (\boldsymbol{\pi}_{t}) - y^i_t \right)
\end{equation}
where the over-dot indicates the time derivative.

\textbf{\emph{2) Choice Stage:}} Once obtained the score, each player is able to map it to a policy  by selecting the greedy action $\arg \max_{a^i} y^i_t(s, a^i)$ at every state. To ensure the map is continuous and single-valued, a smooth and strongly convex regularizer is used to yield the \emph{choice map} $\sigma^i$ from score to policy
\begin{equation*}
  \sigma^i\colon y^i  \to \arg \max_{\pi^i \in \Pi^i} \bigg\{ \sum_{a^i} y^i(s,a^i) \pi^i(s,a^i) - h^i(\pi^i(\cdot | s)) \bigg\}_{s \in \mathcal{S}}.
\end{equation*}
Such $h^i$ is also termed as penalty or smoothing function in some references \citep{coucheney2015penalty, nesterov2005smooth}. Here we consider the (negative) Gibbs entropy, a commonly used form in RL \citep{nachum2017bridging, haarnoja2018softa}
\begin{equation}
\label{eq_entropic_regularizer}
  h^i(\pi^i (\cdot | s)) = \epsilon \sum_{a^i} \pi^i(s, a^i) \log \pi^i(s, a^i)
\end{equation}
which is continuously differentiable and $\epsilon$-strongly convex with respect to $L^1$-norm. $\epsilon > 0$ is known as entropic parameter. A straightforward benefit with entropic regularizer is the closed-form expression of choice map, which is a soft-max function
\begin{equation}
\label{eq_choice_map}
[\sigma^i(y^i)](s,a^i) = \frac{\exp{\left(\frac{1}{\epsilon} y^i(s,a^i) \right)} }{ \sum_{b} \exp{\left( \frac{1}{\epsilon} y^i(s,b) \right) } }.
\end{equation}
When $\epsilon \to 0$, the choice map tends to select the greedy action with the highest score at every state. When $\epsilon$ is arbitrarily large, the policy is like to be uniformly random.

\textbf{\emph{3) Game Stage: }} With the mapped policy $\pi^i_t = \sigma^i (y^i_t), \forall i \in \mathcal{N}$, all players play in the game and observe $\rho_{\boldsymbol{\pi}_t}$ and $A^i_{\boldsymbol{\pi}_t}$ at every state and action. Thus, the learning system in (\ref{eq_yi_dynamics}) operates continuously. For finite MGs, if the game model (reward and transition functions) are known, the exact solutions of $\rho_{\boldsymbol{\pi}_t}$ and $A^i_{\boldsymbol{\pi}_t}$ at given $\boldsymbol{\pi}_t$ can be analytically calculated by linear algebra.

\subsection{Convergence to Nash distribution}

With all players following the scheme as per above, the \emph{continuous-time learning dynamics} of the whole system can be written in stacked form
\begin{equation}
\label{eq_stacked_dynamics}
\tag{CTLD}
  \left\{
  \begin{aligned}
    \dot{\boldsymbol{y}}_t & = \eta \left( \boldsymbol{w} (\boldsymbol{\pi}_{t})  - \boldsymbol{y}_t \right) \\
    \boldsymbol{\pi}_t & = \boldsymbol{\sigma}(\boldsymbol{y}_t)
  \end{aligned} \right.
\end{equation}
where $\boldsymbol{y}_t = ( y^i_t )_{i \in \mathcal{N}}$, $\boldsymbol{w}(\boldsymbol{\pi}_t) = (w^i(\boldsymbol{\pi}_t))_{i \in \mathcal{N}}$, and $\boldsymbol{\sigma}(\boldsymbol{y}_t) = (\sigma^i(y^i_t))_{i \in \mathcal{N}}$. Bounded reward and soft-max choice map make $\boldsymbol{w} \circ \boldsymbol{\sigma}$ a continuous and bounded function. Hence the existence of a fixed point of \ref{eq_stacked_dynamics} is guaranteed by Brouwer's fixed point theorem \citep{gale1979the}. Denote $\bar{\boldsymbol{y}} = (\bar{y}_i)_{i \in \mathcal{N}}$ as the fixed point satisfying $\bar{\boldsymbol{y}} =  \boldsymbol{w} \circ \boldsymbol{\sigma} (\bar{\boldsymbol{y}})$, and let $\bar{\boldsymbol{\pi}} = (\bar{\pi}_i)_{i \in \mathcal{N}}$ be the induced policy profile with $\bar{\boldsymbol{\pi}} = \boldsymbol{\sigma} (\bar{\boldsymbol{y}})$.

\begin{theorem}
  \label{thm:nash_distribution}
1) If $\boldsymbol{\pi}_* = (\pi^i_*)_{i \in \mathcal{N}}$ is a Nash equilibrium to the Markov game with \emph{regularized} payoff, i.e. $U^i (\pi^i, \boldsymbol{\pi}^{-i}) = u^i (\pi^i, \boldsymbol{\pi}^{-i}) - h^i(\pi^i)$ and $U^i ({\pi}^i_*, {\boldsymbol{\pi}}^{-i}_* ) \ge U^i ({\pi}^i, {\boldsymbol{\pi}}^{-i}_* )$, $\forall \pi^i \in \Pi^i, i \in \mathcal{N}$,
then ${\boldsymbol{y}}_* = \boldsymbol{w}({\boldsymbol{\pi}}_*)$ is the fixed point of \ref{eq_stacked_dynamics}.

\setlength\parindent{12pt} 2) The converse is true if each player's original payoff $u^i$ is individually concave in the sense that $u^i(\pi^i, \boldsymbol{\pi}^{-i})$ is concave in $\pi^i$ for all $\boldsymbol{\pi}^{-i} \in \boldsymbol{\Pi}^{-i}$, $\forall i \in \mathcal{N}$.
\end{theorem}

Note that in Theorem \ref{thm:nash_distribution}, the equilibrium is modified to take into account the influence of regularizer. It is sometimes referred to as \emph{Nash distribution} \citep{leslie2005ind, gao2021on} to distinguish from the Nash equilibrium with original payoff. When the entropic regularizer in (\ref{eq_entropic_regularizer}) takes $\epsilon \to 0$, the Nash distribution coincides with the Nash equilibrium.
One condition for the global equivalence between fixed-point policy and Nash distribution is the individual concavity of game payoff. In many scenarios \citep{Schofield2002LocalNE, rat2013chara, maz2019on}, a local Nash is sometimes easier to use than an expensive global solution.
The following corollary extends the second part of Theorem \ref{thm:nash_distribution} to local cases by restricting the interested domain to a neighbor of the fixed point.

\begin{corollary}
  \label{corl:local_nash_distribution}
  Let $\bar{\boldsymbol{y}}$ be the fixed point of $\boldsymbol{w} \circ \boldsymbol{\sigma}$. If $u^i$ is locally individually concave around $\bar{\boldsymbol{\pi}} = \boldsymbol{\sigma} (\bar{\boldsymbol{y}})$ for all players, then $\bar{\boldsymbol{\pi}} $ is a \emph{local} Nash distribution.
\end{corollary}

Now we analyze the convergence property of \ref{eq_stacked_dynamics} based on the Lyapunov stability theory of dynamical systems \citep{khalil2002nonlinear}. Consider the Fenchel-coupling function \citep{mer2019learning} and by summing over all states, define
\begin{equation*}
\begin{aligned}
  F^i (\pi^i, y^i) = \max_{\pi \in \Pi^i} \sum_s \Big( \textstyle \sum_{a^i} y^i(s,a^i) \pi(s,a^i) - h^i(\pi(\cdot | s)) \Big)
  \\ - \sum_s  \Big( \textstyle \sum_{a^i} y^i(s,a^i) \pi^i(s,a^i) - h^i(\pi^i(\cdot | s)) \Big)
\end{aligned}
\end{equation*}
for any $(\pi^i, y^i)$ pair. Naturally $F^i (\pi^i, y^i) \ge 0$, and with entropic $h^i$ and soft-max $\sigma^i$ as per above, it is continuously differentiable. By staying at the fixed-point policy $\bar{\boldsymbol{\pi}}$, we can take $\sum_i F^i (\bar{\pi}^i, {y}^i_t)$ as the Lyapunov function and calculate its time derivative along the solution of \ref{eq_stacked_dynamics}. Before presenting the convergence theorem, we introduce the following definition to characterize the property of Markov games. For ease of notation and analysis, functions over state and action sets are considered as matrices of the size $|\mathcal{S}| \times |\mathcal{A}^i|$. Let $\langle \cdot, \cdot \rangle$ be the Frobenius inner product for the sum of the component-wise product of two matrices, and $\| \cdot \|$ be the induced matrix norm with $\| \pi\|^2 = \sum_{s} \| \pi(\cdot | s)\|^2$. For $n$-player aggregation, the above two notations indicate the sum over all $i \in \mathcal{N}$.

\begin{definition}[Monotonicity and hypomonotonicity]
\label{defn:mu_monotone}
A Markov game is called \emph{monotone} if for any policy profiles $\boldsymbol{\pi}$ and $\boldsymbol{\pi}_{\dagger}$, it has
$\langle \boldsymbol{w}(\boldsymbol{\pi}) - \boldsymbol{w} (\boldsymbol{\pi}_{\dagger}) , \boldsymbol{\pi} - \boldsymbol{\pi}_{\dagger}  \rangle \le 0$. If the inequality holds only for $\langle \boldsymbol{w}(\boldsymbol{\pi}) - \boldsymbol{w} (\boldsymbol{\pi}_{\dagger}) , \boldsymbol{\pi} - \boldsymbol{\pi}_{\dagger}  \rangle \le \mu \| \boldsymbol{\pi} - \boldsymbol{\pi}_{\dagger} \|^2$ with some $\mu > 0$, the game is called \emph{$\mu$-hypomonotone}.
\end{definition}

\begin{theorem}
  \label{thm:convergence}
  Consider the Markov game and the learning scheme provided in \ref{eq_stacked_dynamics}. Assume there are a finite number of isolated fixed points $\bar{\boldsymbol{y}}$ of $\boldsymbol{w} \circ \boldsymbol{\sigma}$.
  If the game is $\mu$-hypomonotone ($\mu \ge 0$) and the entropic regularizer chooses $\epsilon > 2\mu$, then

  \setlength\parindent{12pt} 1) players' scores $\boldsymbol{y}_t$ converge to a fixed point $\bar{\boldsymbol{y}}$.

  \setlength\parindent{12pt} 2) If further the game is individually concave, players' policies $\boldsymbol{\pi}_t$ converge to a Nash distribution $\bar{\boldsymbol{\pi}} = \boldsymbol{\sigma} (\bar{\boldsymbol{y}})$.

  \setlength\parindent{12pt} 3) If instead the game is only locally individually concave around $\bar{\boldsymbol{\pi}} = \boldsymbol{\sigma} (\bar{\boldsymbol{y}})$, players' policies $\boldsymbol{\pi}_t$ converge to a local Nash distribution $\bar{\boldsymbol{\pi}}$.
  %\begin{enumerate}
%    \item [i)] players' scores $\boldsymbol{y}_t = (y^i_t)_{i \in \mathcal{N}}$ converge to a fixed point $\bar{\boldsymbol{y}}$.
%    \item [ii)] If further the game is individually concave, players' policies $\boldsymbol{\pi}_t = (\pi^i_t)_{i \in \mathcal{N}}$, $ {\pi}^i_t ={\sigma} ({y}^i_t)$ converge to a Nash distribution $\bar{\boldsymbol{\pi}} = \boldsymbol{\sigma} (\bar{\boldsymbol{y}})$.
%    \item [iii)] If the game is only locally individually concave around $\bar{\boldsymbol{\pi}} = \boldsymbol{\sigma} (\bar{\boldsymbol{y}})$, players' policies $\boldsymbol{\pi}_t$ converge to a local Nash distribution $\bar{\boldsymbol{\pi}}$.
%  \end{enumerate}
\end{theorem}

We here specify Gibbs entropy as the regularizer function in the operation of \ref{eq_stacked_dynamics}. In fact, there are many other forms of regularizers, like Tsallis entropy and Burg entropy \citep{coucheney2015penalty}. A generalization of convergence theorem with arbitrary regularizers is given in Appendices, provided that the regularizer functions meet certain conditions.

In the above theorem, monotonicity is considered as a special case of hypomonotonicity with $\mu = 0$. If the game is monotone, players are able to converge to Nash equilibrium by taking arbitrarily small $\epsilon$. When $\mu>0$, to ensure convergence, the system has to choose large enough $\epsilon > 2\mu$ to compensate the shortage of monotonicity. But too large $\epsilon$ deviates Nash distribution away from Nash equilibrium, so a tradeoff exists \citep{gao2021on}.

\begin{proposition}
\label{prop:monotone}
  For any MGs there always exists a finite $\mu \ge 0$ such that $\langle \boldsymbol{w}(\boldsymbol{\pi}) - \boldsymbol{w} (\boldsymbol{\pi}_{\dagger}) , \boldsymbol{\pi} - \boldsymbol{\pi}_{\dagger}  \rangle \le \mu \| \boldsymbol{\pi} - \boldsymbol{\pi}_{\dagger} \|^2$ holds for any two policy profiles.
\end{proposition}

\section{Empirical policy optimization}

Application of CTLD to practical large games faces obstacles from two aspects: 1) it is computationally expensive, if not impossible, to analytically evaluate players' policies on large state/action sets; 2) policies in large-scale problems are not explicitly expressed but are parameterized by approximators like NNs \citep{mnih2015human, vinyals2019grandmaster}. In this section, we develop an \emph{empirical policy optimization} (EPO) algorithm to learn parameterized policies via reinforcement learning.

We first transform CTLD to a \emph{discrete-time learning dynamics} (DTLD) based on stochastic approximation \citep{benaim1999dyna}. The evolution of all players follows the discrete-time update rule
\begin{equation}
\label{eq_discrete_time_dynamics}
\tag{DTLD}
\left\{
\begin{aligned}
  y^i_{l+1} = & y^i_l + \alpha_l \eta ( \hat{w}^i_l - y^i_l  )  \\
  \pi^i_{l+1} = & \sigma^i (y^i_{l+1})
\end{aligned}
\right.
\end{equation}
where $l$ indicates the discrete-time iteration, $\hat{w}^i_l$ is the observed (noisy) weighted advantage of $\boldsymbol{\pi}_l$, and $\alpha_l$ is the update step. According to the stochastic approximation theory, the long-term behavior of \ref{eq_discrete_time_dynamics} is related to that of solution trajectories of its mean-field ordinary differential equation, which can coincide with \ref{eq_stacked_dynamics} under certain conditions.

\begin{theorem}
  \label{thm:discrete_convergence}
  Consider the Markov game and the learning scheme provided in \ref{eq_discrete_time_dynamics}. Assume there are a finite number of isolated fixed points $\bar{\boldsymbol{y}}$ of $\boldsymbol{w} \circ \boldsymbol{\sigma}$.
  At every iteration $l$, each player's $\hat{w}^i_l$ is an unbiased estimate of $w^i(\boldsymbol{\pi}_l)$, i.e. $\mathbb{E} [\hat{w}^i_l] = w^i(\boldsymbol{\pi}_l)$, and has $\mathbb{E} [ \| \hat{w}^i_l - w^i(\boldsymbol{\pi}_l) \|^2] \le C$, for some $C \ge 0$. $\| y^i_l\|$ is always finite during the learning process. $\{ \alpha_l\}$ is a deterministic sequence satisfying $\sum_{l=0}^{\infty} \alpha_l = \infty$ and $\sum_{l=0}^{\infty} \alpha_l^2 < \infty$.
  If the game is $\mu$-hypomonotone ($\mu \ge 0$) and the entropic regularizer chooses $\epsilon > 2\mu$, then players' scores $\boldsymbol{y}_l$ converge almost surely to a fixed point $\bar{\boldsymbol{y}}$.
\end{theorem}

Under Theorem \ref{thm:discrete_convergence}, the almost sure convergence of $\boldsymbol{\pi}_l$ to a (local) Nash distribution follows the proof of Theorem \ref{thm:convergence} under the (local) individual concavity of game payoff.

In large games, assume each player defines a policy network $\hat{\pi}^i$, parameterized by $\theta^i$. The choice map $\sigma^i$ with input $y^i_l$ becomes finding a group of parameters $\theta^i_{l}$ that minimize the loss
\begin{equation*}
\label{eq_policy_loss}
 \min_{\theta^i}  \mathcal{L}^i (\theta^i) = \min_{\theta^i} \sum_{s,a^i} y^i_l (s,a^i) \hat{\pi}^i(s,a^i | \theta^i) - h^i (\hat{\pi}^i(\theta^i)).
\end{equation*}
To avoid extreme change of policy behaviors $\{\hat{\pi}^i_l \}$ along iterations, we restrict the new $\hat{\pi}^i (\theta^i_l)$ is trained along the loss gradient $\partial \mathcal{L}^i / \partial \theta^i$, starting from last $\theta^i_{l-1}$, and borrow the idea from \citep{Schulman2017ProximalPO} to use an early stop to bound the $\mathrm{KL}$ divergence between the new and old policies, i.e. $\mathbb{E}_{s \sim \rho_{l-1}} [D_{\mathrm{KL}} (\hat{\pi}^i(\theta^i_{l-1}) \| \hat{\pi}^i(\theta^i_{l}))] \le c$.

If we specify DTLD with $\alpha_l = \frac{1}{l}$, $\eta = 1$ and ignore the noise effect, $y^i_l$ is actually the average of past weighted advantages
\begin{align}
\label{eq_average_score}
  y^i_l (s,a^i) = & \frac{1}{l} \sum_{j=0}^{l-1} \rho_j (s) A^i_j (s, a^i) \notag \\
  = & \frac{1}{l} \sum_{j=0}^{l-1} \rho_j(s) Q^i_j(s,a^i) - \frac{1}{l} \sum_{j=0}^{l-1} \rho_j(s) V^i_j(s).
\end{align}
Because state-dependent terms make no difference to the gradient $\partial \mathcal{L}^i / \partial \theta^i$, the above sum of values can be replaced by an \emph{empirical} value network $\hat{V}^i(s|\phi^i)$ that learns the average of historical weighted values, i.e. $\hat{V}^i(s|\phi^i) \approx \frac{1}{\sum_{j=0}^{l-1} \rho_j(s)} \sum_{j=0}^{l-1} \rho_j(s) V^i_j(s)$. The weighted calculation $ \rho_j(s) [\dots ]$ is equivalent to the expectation $\frac{1}{1-\gamma} \mathbb{E}_{s \sim \rho_j} [\dots]$, and can be further approximated by samples observed at every iteration \citep{schulman2015trust}.

With the value network, the score becomes
$
y^i_l (s,a^i) = \frac{1}{l} \sum_{j=0}^{l-1} \rho_j(s) \big( Q^i_j(s,a^i) - \hat{V}^i(s|\phi^i) \big)
$.
The return $G^i_{j,k}$ on the on-policy trajectory $(s_k, \boldsymbol{a}_k, s_{k+1}, \boldsymbol{a}_{k+1}, \dots)$ generated by $\hat{\boldsymbol{\pi}}_j$ is an unbiased estimate of $Q^i_j(s_k, a^i_k)$, but suffers from high variance. A commonly used form in modern RL is the Generalized Advantage Estimator (GAE) \citep{schulman2016high}, which is a biased and low-variance estimate. For any segment of trajectory $(s_k, a^i_k, r^i_{k}, s_{k+1}, a^i_{k+1}, r^i_{k+1}, \dots ) $ in the historical experience, with the support of $\hat{V}^i(\phi^i)$, player $i$'s $\lambda$-GAE is defined as
$
  \hat{A}^i_{k} = \sum_{\nu =0}^{\infty} (\gamma \lambda)^{\nu}  \delta^i_{k + \nu}
$,
where $\delta^i_{k + \nu} = r^i_{k + \nu} + \gamma \hat{V}^i(s_{k + \nu+1} |\phi^i) - \hat{V}^i(s_{k + \nu} |\phi^i) $ is the temporal difference, and $\lambda  \in [0, 1]$ is a constant that balances the bias and variance of estimate.
The policy loss now becomes
%\begin{equation}
%\label{eq_policy_loss_sample}
%      \mathcal{L}^i (\theta^i) = \sum_{(s_k, a^i_k) \in \mathcal{D}^i} \left[  \hat{A}^i (s_k, a_k) \hat{\pi}^i(s_k, a^i_k | \theta^i) - h^i (\hat{\pi}^i(s_k | \theta^i)\right]
%\end{equation}
\begin{equation*}
\label{eq_policy_loss_sample}
      \mathcal{L}^i (\theta^i) = \sum_{\mathcal{D}^i} [  \hat{A}^i (s_k, a^i_k) \hat{\pi}^i(s_k, a^i_k | \theta^i) - h^i (\hat{\pi}^i(s_k | \theta^i) ]
\end{equation*}
where $\mathcal{D}^i$ is the experience of player $i$ in the entire history.
Empirically, the clipping technique proposed by \citet{Schulman2017ProximalPO} is helpful to stabilize the optimization.

Algorithm \ref{algm:epo} summarizes the whole process of EPO. The learning is totally distributed in the sense that each player trains its value and policy networks based on own observations of states, actions, and rewards (Lines 5\textasciitilde 9). It requires no knowledge of game structure (how many players are playing and what the others' rewards are defined) and does not need to monitor the other behaviors. All players play their current policies (Line 3) in the same game. While another $n$-player framework--PSRO \citep{lanctot2017a} has to match each player with specific opponents.

The procedure of EPO shows similarity to that of Proximal Policy Optimization (PPO) \citep{Schulman2017ProximalPO}, but with fundamental difference. EPO follows the idea of CTLD that updates multiple players' policies based on the aggregation of their in-game performance over the past iterations, in contrast to PPO that optimizes single-agent policy with only experience of the current iteration. With the whole historical experience, EPO updates multi-player policies in the direction of Nash equilibrium.

\begin{algorithm}[h]
\caption{Empirical Policy Optimization for $n$-Player Markov Games}
\begin{algorithmic}[1]
\STATE Initialize policy and value parameters, $\boldsymbol{\theta}_0 = (\theta^i_0)_{i \in \mathcal{N}}$, $\boldsymbol{\phi}_0 = (\phi^i_0)_{i \in \mathcal{N}}$; define experience buffer $\mathcal{D}^i = \emptyset$, $\forall i \in \mathcal{N}$;
select entropy parameter $\epsilon$, GAE parameter $\lambda $, KL divergence threshold $c$;
\FOR{$l = 0, 1, \dots, $}
\STATE  Players play their own $\hat{\pi}^i(\theta^i_l)$ in game and observe trajectories $\tau^i_l = \{ (s_k, {a}^i_k, {r}^i_k, s_{k+1})\}$;
  \FOR{each player $i$}
    \STATE Calculate return $G^i_k $ along $\tau^i_l$ and store $\{ (s_k, {a}^i_k, {r}^i_k, s_{k+1}, G^i_k)\}$ in $\mathcal{D}^i$;
    \STATE Fit empirical value network $\hat{V}^i (\phi^i)$ over $\mathcal{D}^i$ by regression on mean-squared error
    $
      \min_{\phi^i} \sum_{s_k \in \mathcal{D}^i} \big( \hat{V}^i (s_k|\phi^i) - G^i_k \big)^2
    $;
    \STATE Compute $\lambda$-GAE $\hat{A}^i_{k}$ for every sample in $\mathcal{D}^i$ based on the regressed $\hat{V}^i (\phi^i)$;
    \STATE Train policy network along the gradient of policy loss $\mathcal{L}^i (\theta^i)$, starting from the current $\theta^i_l$ with KL divergence threshold $\mathbb{E}_{s_k \sim \tau^i_l} [D_{\mathrm{KL}} (\hat{\pi}^i(\theta^i_{l}) \| \hat{\pi}^i(\theta^i))] \le c$;
    \STATE Take the trained $\theta^i$ as $\theta^i_{l+1}$.
  \ENDFOR
\ENDFOR
\end{algorithmic}
\label{algm:epo}
\end{algorithm}

\section{Experiments}

In experiments, we consider 2-player \emph{Soccer} game \citep{littman1994markov, zhu2020online}, 3-player \emph{Cournot-Competition} game \citep{mer2019learning}, and 2-player \emph{Wimblepong} game\footnote{\url{https://github.com/aalto-intelligent-robotics/wimblepong}}.

\subsection{Numerical examples}

We apply the proposed CTLD to learn Nash equilibria for the first two games. For comparison, we consider Iterated Best Response (IBR) \citep{nar2007using}, Fictitious Play (FP) \citep{heinrich2015fictitious}, Policy Space Response Oracle (PSRO) \citep{lanctot2017a}, and Exploitability Descent (ED) in tabular form \citep{lockhart2019comp}, and use policy iteration \citep{sutton2018reinforcement} as their oracles for best response. PSRO relies on a meta-solver to synthesize meta-strategies for each player, so we choose linear programming \citep{raghavan1994zero} for 2-player case and the EXP-D-RL method proposed in \citep{gao2021on} for 3-player case. ED is originally proposed in \citep{lockhart2019comp} for two-player zero-sum game, but here is also applied to the 3-player game. Hyperparameters have been empirically selected, and detailed implementation is presented in Appendices.

\begin{figure}
    \centering
    \begin{subfigure}[t]{0.22\textwidth}
        \centering
        \includegraphics[width=1\textwidth, height=0.65\textwidth]{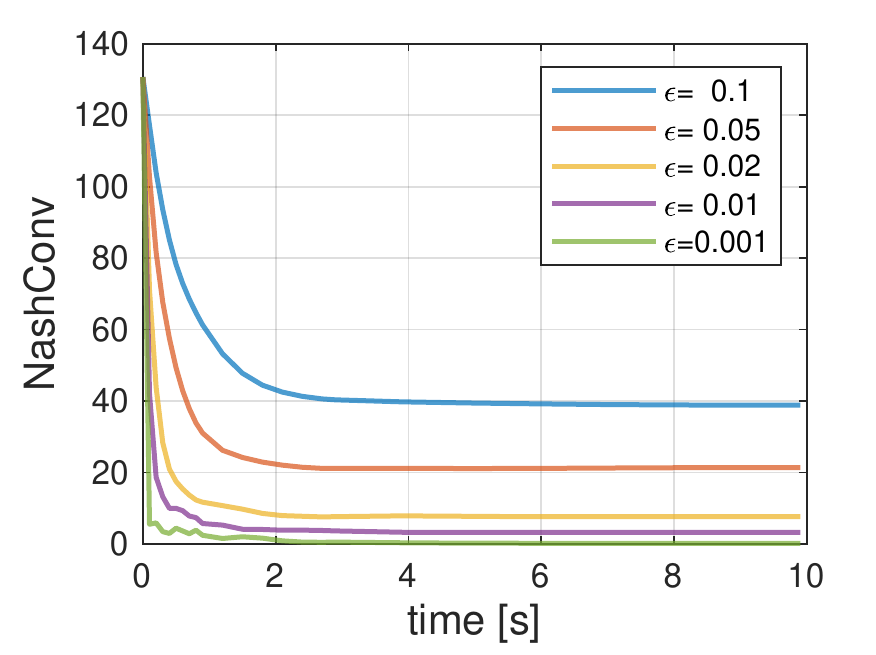}
        \caption{CTLD}
    \end{subfigure}
    \begin{subfigure}[t]{0.22\textwidth}
        \centering
        \includegraphics[width=1\textwidth, height=0.65\textwidth]{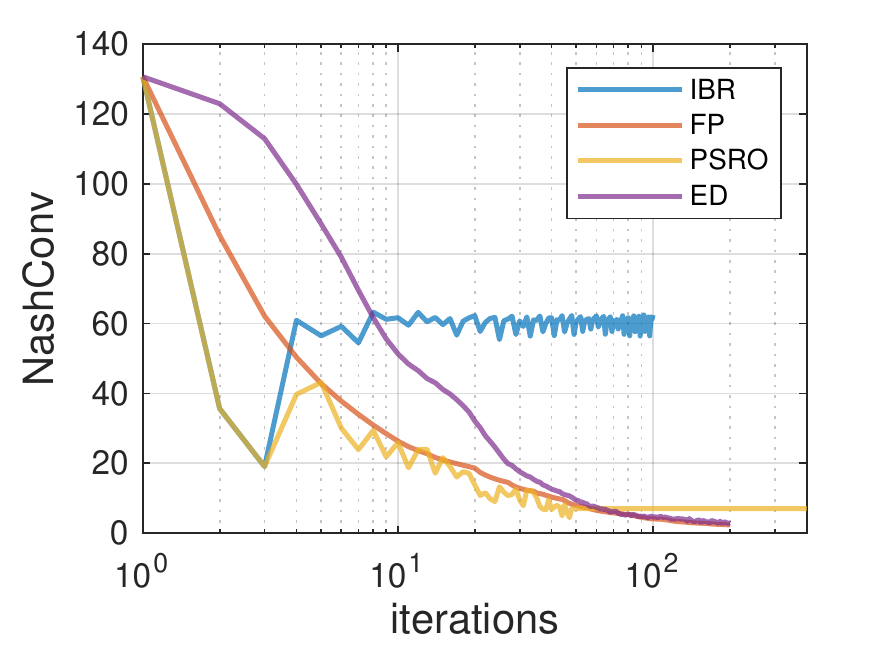}
        \caption{IBR vs FP vs PSRO vs ED}
    \end{subfigure}
    \caption{$\mathrm{NashConv}$ learning curves on Soccer.}
    \label{fig_soccer_learning}
\end{figure}

\begin{figure}
    \centering
    \vspace{1.1em}
    \begin{subfigure}[t]{0.22\textwidth}
        \centering
        \includegraphics[width=1\textwidth, height=0.65\textwidth]{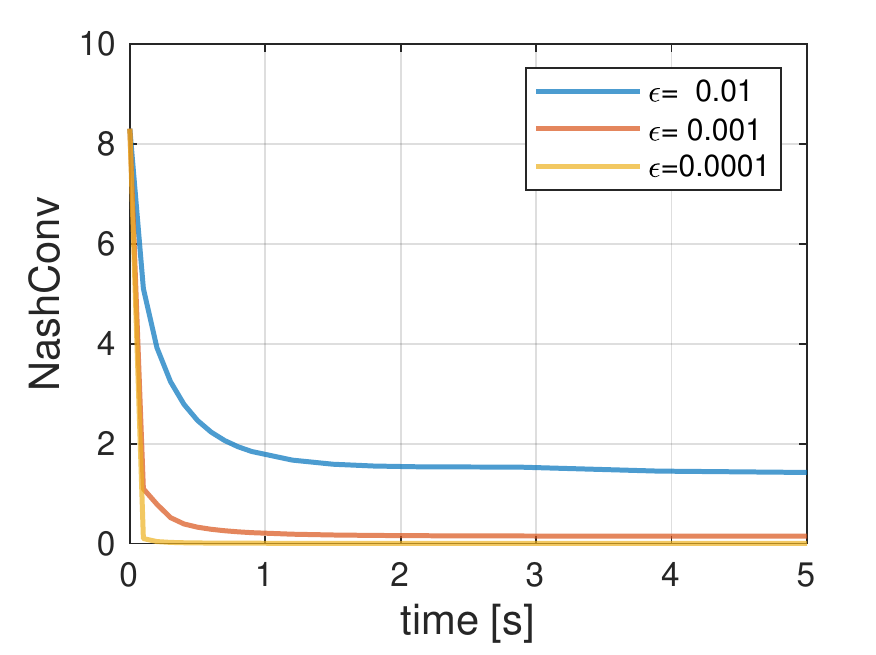}
        \caption{CTLD}
    \end{subfigure}
    \begin{subfigure}[t]{0.22\textwidth}
        \centering
        \includegraphics[width=1\textwidth, height=0.65\textwidth]{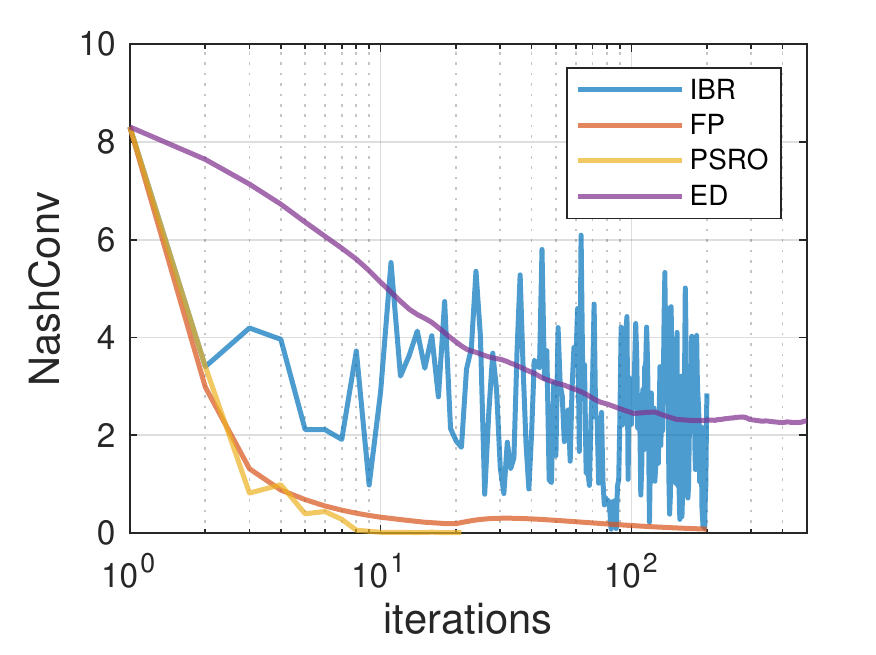}
        \caption{IBR vs FP vs PSRO vs ED}
    \end{subfigure}
    \caption{$\mathrm{NashConv}$ learning curves on Cournot Competition.}
    \label{fig_tfcc_learning}
\end{figure}

The $\mathrm{NashConvs}$ of each method along the learning process are plotted in Figure \ref{fig_soccer_learning} and Figure \ref{fig_tfcc_learning}.
Since CTLD and the other methods are running in different time scales, their results are presented separately in different plots. In both experiments, CTLD remains convergent under any regularizer parameter $\epsilon$, and is able to approach Nash equilibria with arbitrary precision if $\epsilon$ is close enough to 0.
Another empirical learning method, FP, also shows consistent convergence property. PSRO is remarkable in approaching Nash equilibrium in Cournot Competition, but ends up with a noticeable $\mathrm{NashConv}$ gap in Soccer game. The convergence of ED in 2-player case is guaranteed by the theoretical results in \citep{lockhart2019comp}, but the argument is not valid for $n$-player games with $n > 2$, resulting in a large gap of $\mathrm{NashConv}$ in Cournot Competition. IBR suffers from strategic cycles, so it is hard to converge.

\begin{figure}
    \centering
    \begin{subfigure}[t]{0.22\textwidth}
        \centering
        \includegraphics[width=1\textwidth, height=0.5\textwidth]{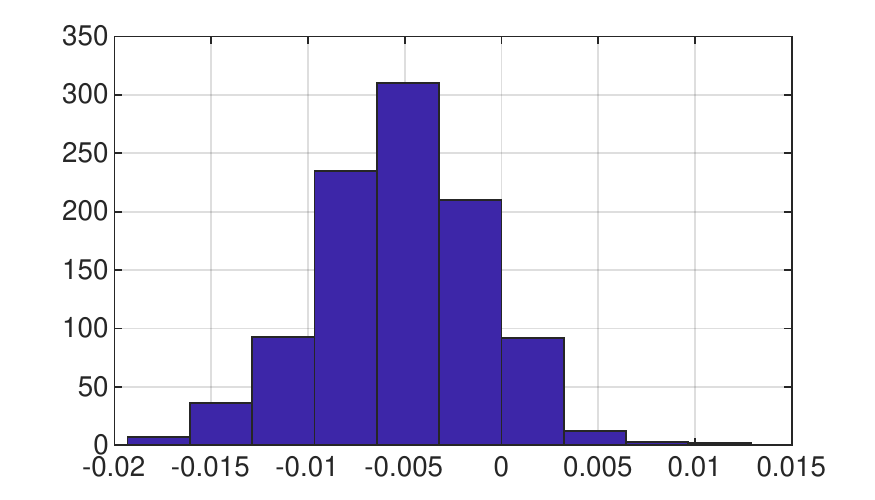}
        \caption{Soccer game}
    \end{subfigure}
    \begin{subfigure}[t]{0.22\textwidth}
        \centering
        \includegraphics[width=1\textwidth, height=0.5\textwidth]{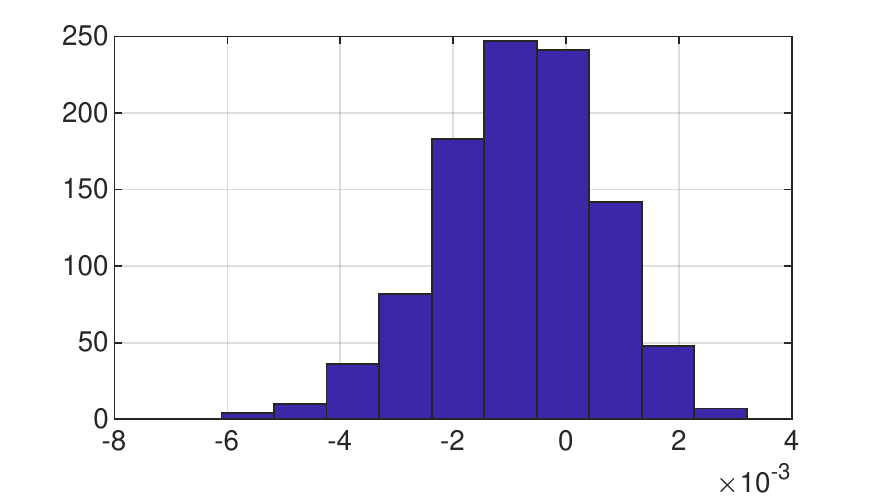}
        \caption{Cournot Competition}
    \end{subfigure}
    \caption{Histograms of sampled $ \| \boldsymbol{w}(\boldsymbol{\pi}) - \boldsymbol{w}(\boldsymbol{\pi_{\dagger}}) \| / \| \boldsymbol{\pi} - \boldsymbol{\pi_{\dagger}} \|^2 $.}
    \label{fig_sampled_mu}
\end{figure}

We also numerically investigate the hypomonotone values $\mu$ of two games.
By randomly choosing two policy profiles $\boldsymbol{\pi}$ and $\boldsymbol{\pi_{\dagger}}$, the result of $ \| \boldsymbol{w}(\boldsymbol{\pi}) - \boldsymbol{w}(\boldsymbol{\pi_{\dagger}}) \| / \| \boldsymbol{\pi} - \boldsymbol{\pi_{\dagger}} \|^2 $ is an under-estimate of true $\mu$. The distributions of 1000 samples in two games are plotted in Figure \ref{fig_sampled_mu}. The true $\mu$ is inferred to be greater than 0.0129 in Soccer and greater than 0.0032 in Cournot Competition. It reflects that the convergence condition $\epsilon > 2\mu$ in Theorem \ref{thm:convergence} is not that strict, since we have observed with smaller $\epsilon <  2 \mu$, the CTLD still converges in both games.

Another hyperparameter in CTLD is the learning rate $\eta$. Additional experiments on the effect of $\eta$ show that large $\eta$ has no influence on the converged results, but is able to accelerate the convergence rate.

\subsection{Large-scale example}

\begin{figure}
  \centering
  \begin{subfigure}[b]{.23\textwidth}
  \centering
  \includegraphics[width=1\textwidth]{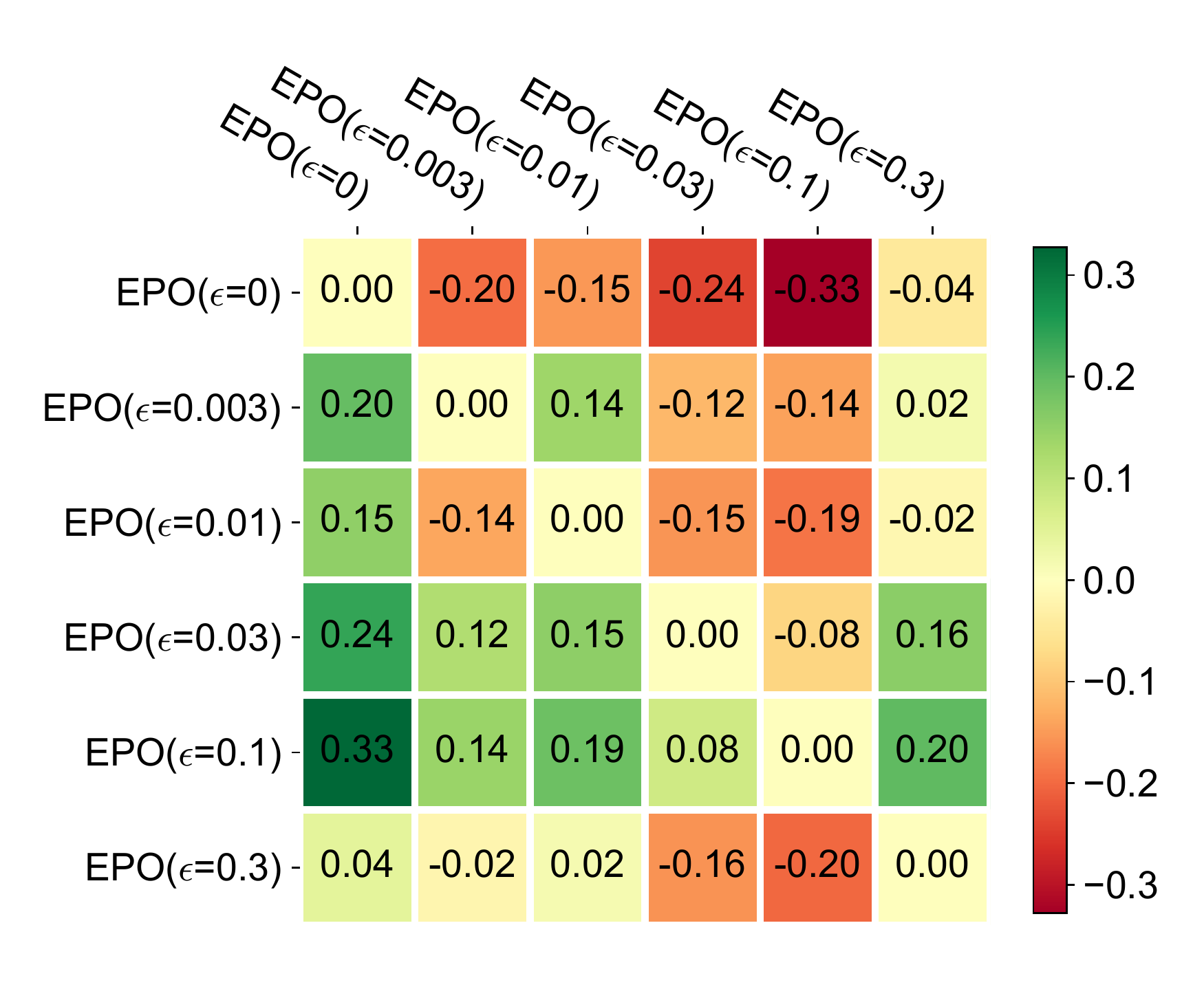}
  \caption{Payoff table}
  \end{subfigure}
  \begin{subfigure}[b]{.23\textwidth}
  \centering
  \includegraphics[width=1\textwidth]{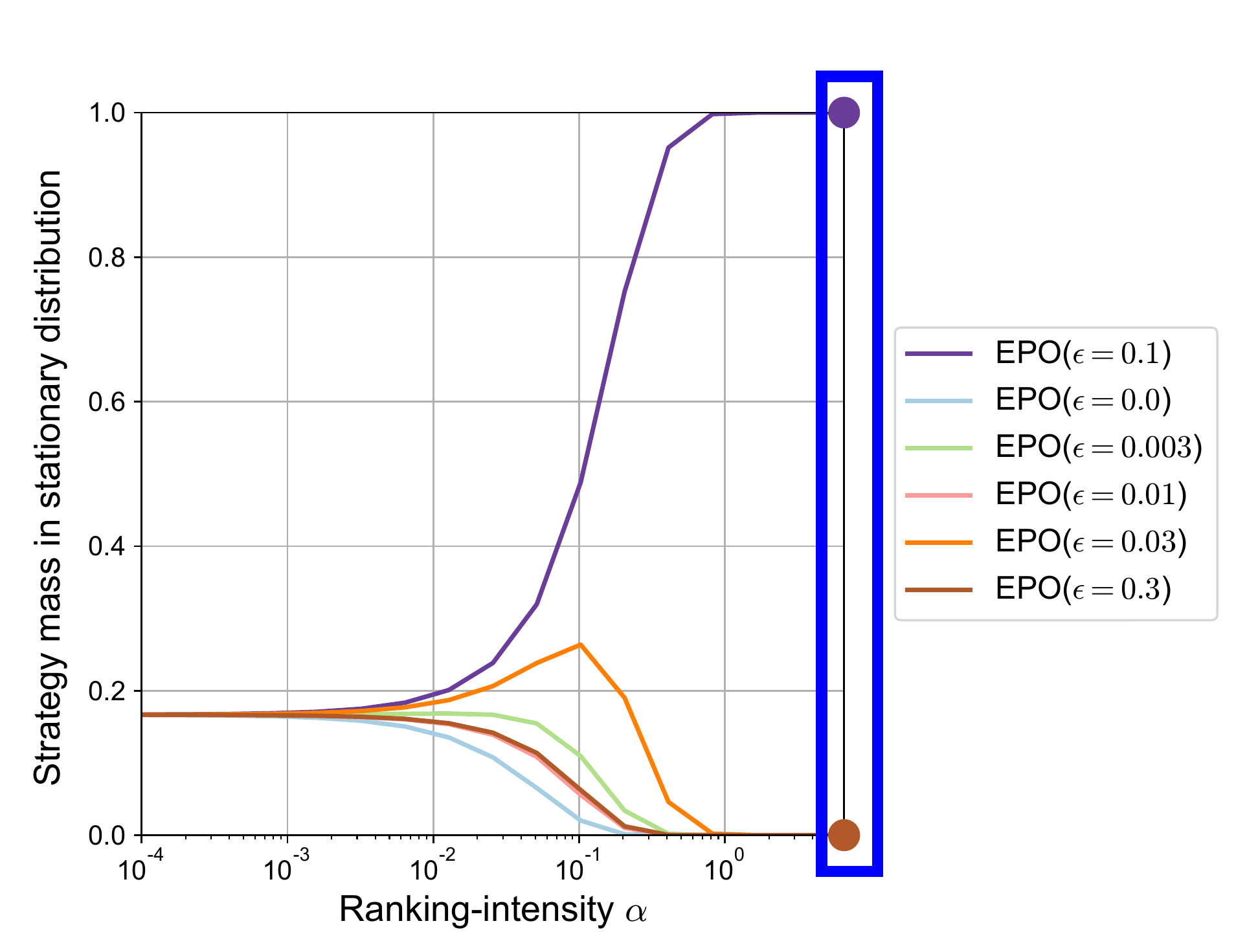}
  \caption{Ranking-intensity sweep}
  \end{subfigure}
  \caption{Evaluation of EPO agents learned under different $\epsilon$. }
  \label{fig_wimblepong_epo_arank}
\end{figure}

The third Wimblepong game is large-scale, so EPO is applied. We run the experiments with different regularizer parameters $\epsilon$ and select common values in RL literature for the rest algorithm parameters. To reduce random errors, each experiment is repeated three times. After 400 iterations, the learned agents under different $\epsilon$ are matched in pairs to evaluate their agent-level payoff table. The payoff value is calculated by the difference between two-side win rates, and is averaged over matches played by agents that are obtained in different runs.
We use the multi-agent evaluation and ranking metric, $\alpha$-Rank \citep{omidshafiei2019alpha}, to evaluate agent rankings, and present the results in Figure \ref{fig_wimblepong_epo_arank}. EPO with $\epsilon=0.1$ shows dominance in playing against the other EPO agents. Small $\epsilon$ causes algorithm to prematurely stop exploration and fall into local optima, while large $\epsilon$ disturbs action selection.

For comparison, we choose Self-Play (SP), Neural Fictitious Self-Play (NFSP) \citep{Heinrich2016DeepRL}, Nash-based PSRO \citep{lanctot2017a}, and PPO \citep{Schulman2017ProximalPO} against a script-based SimpleAI opponent. For fairness, the RL parts of SP, NFSP, and PSRO are all based on a PPO agent. The fictitious player in NFSP is trained by supervised learning based on the historical behavior of fellow agent. The opponent meta-strategy in PSRO is the Nash mixture of historical policies.
The algorithms choose the same parameters as EPO and vary the entropic parameter $\epsilon$ in training objectives to produce a variety of agents. We take the learning process of EPO with the best $\epsilon=0.1$ as baseline and evaluate the relative performance of these algorithms to EPO along the same number of iterations.

\begin{figure}
  \centering
  \begin{subfigure}[b]{.22\textwidth}
  \centering
  \includegraphics[width=1\textwidth, height=0.7\textwidth]{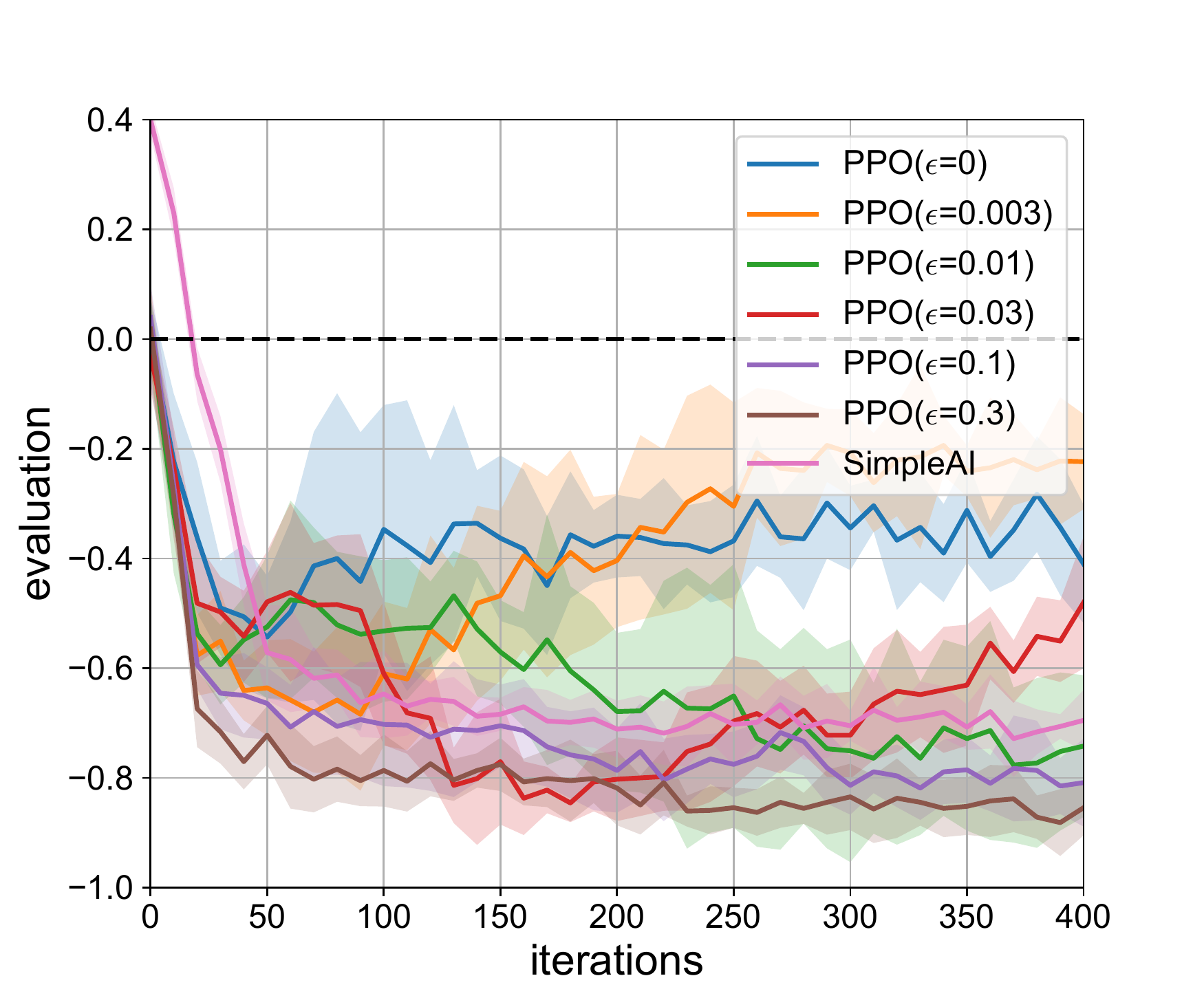}
  \caption{PPO/SimpleAI vs EPO}
  \end{subfigure}
  \begin{subfigure}[b]{.22\textwidth}
  \centering
  \includegraphics[width=1\textwidth, height=0.7\textwidth]{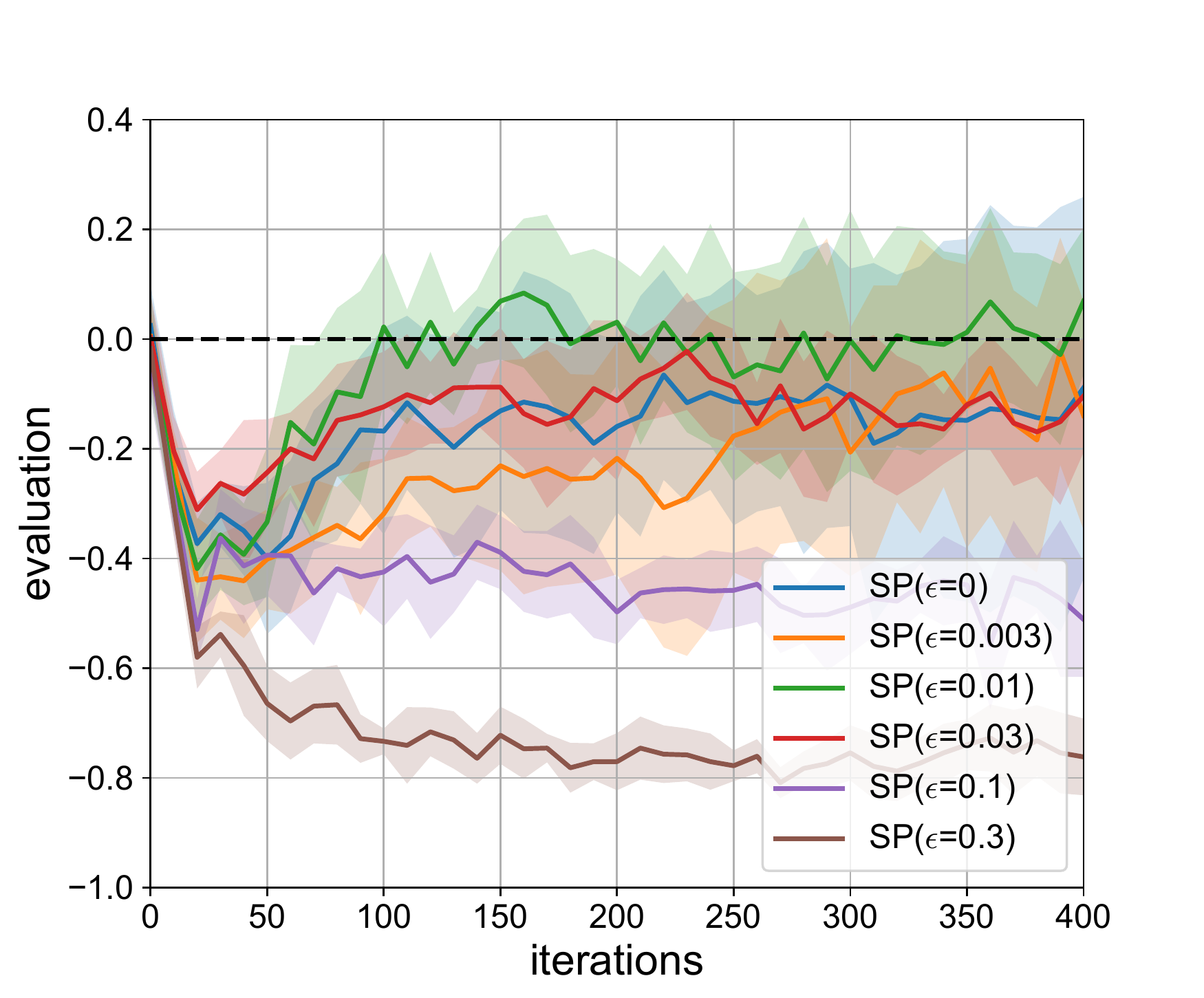}
  \caption{SP vs EPO}
  \end{subfigure}
  \begin{subfigure}[b]{.22\textwidth}
  \centering
  \includegraphics[width=1\textwidth, height=0.7\textwidth]{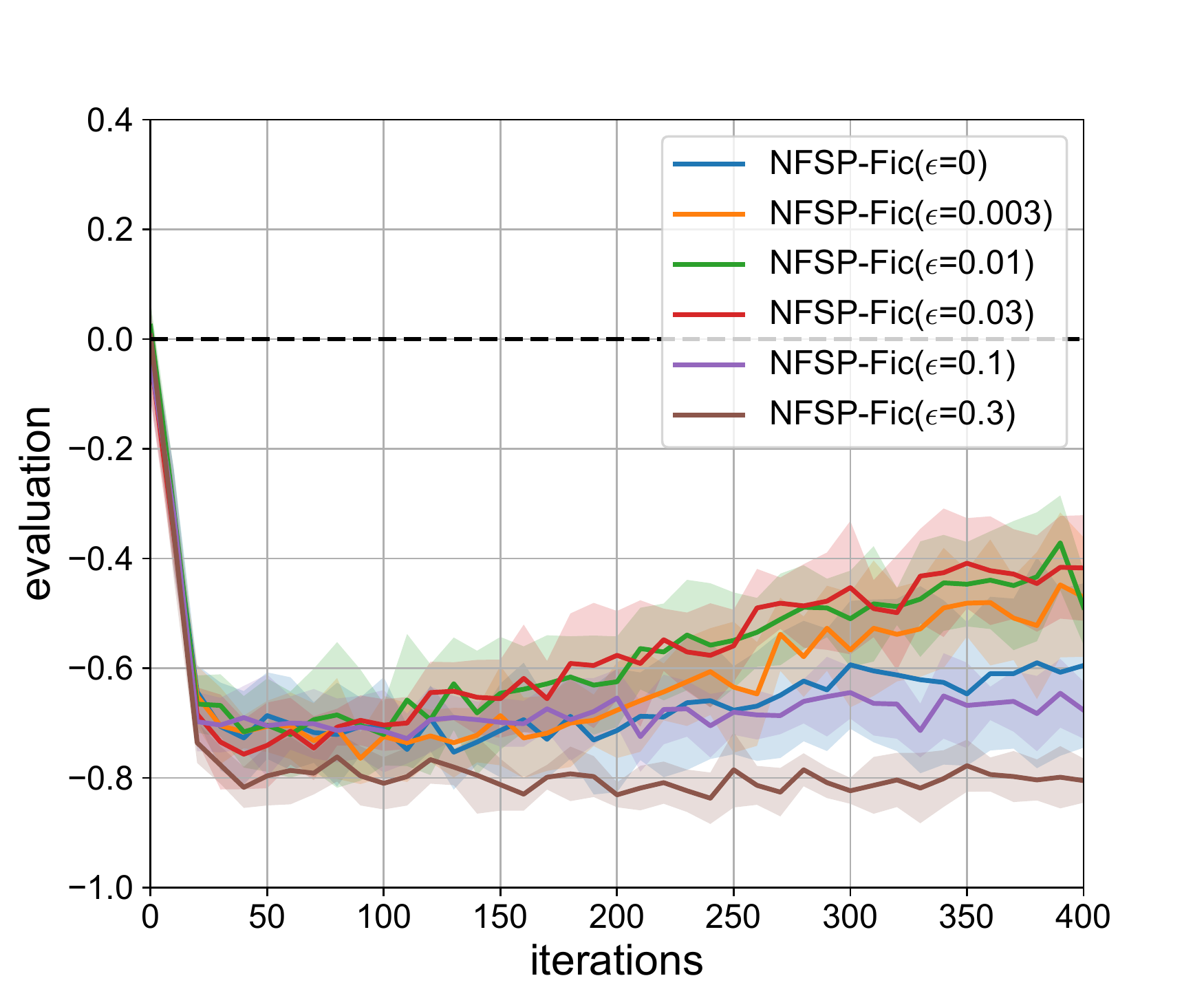}
  \caption{NFSP-Fic vs EPO}
  \end{subfigure}
  \begin{subfigure}[b]{.22\textwidth}
  \centering
  \includegraphics[width=1\textwidth, height=0.7\textwidth]{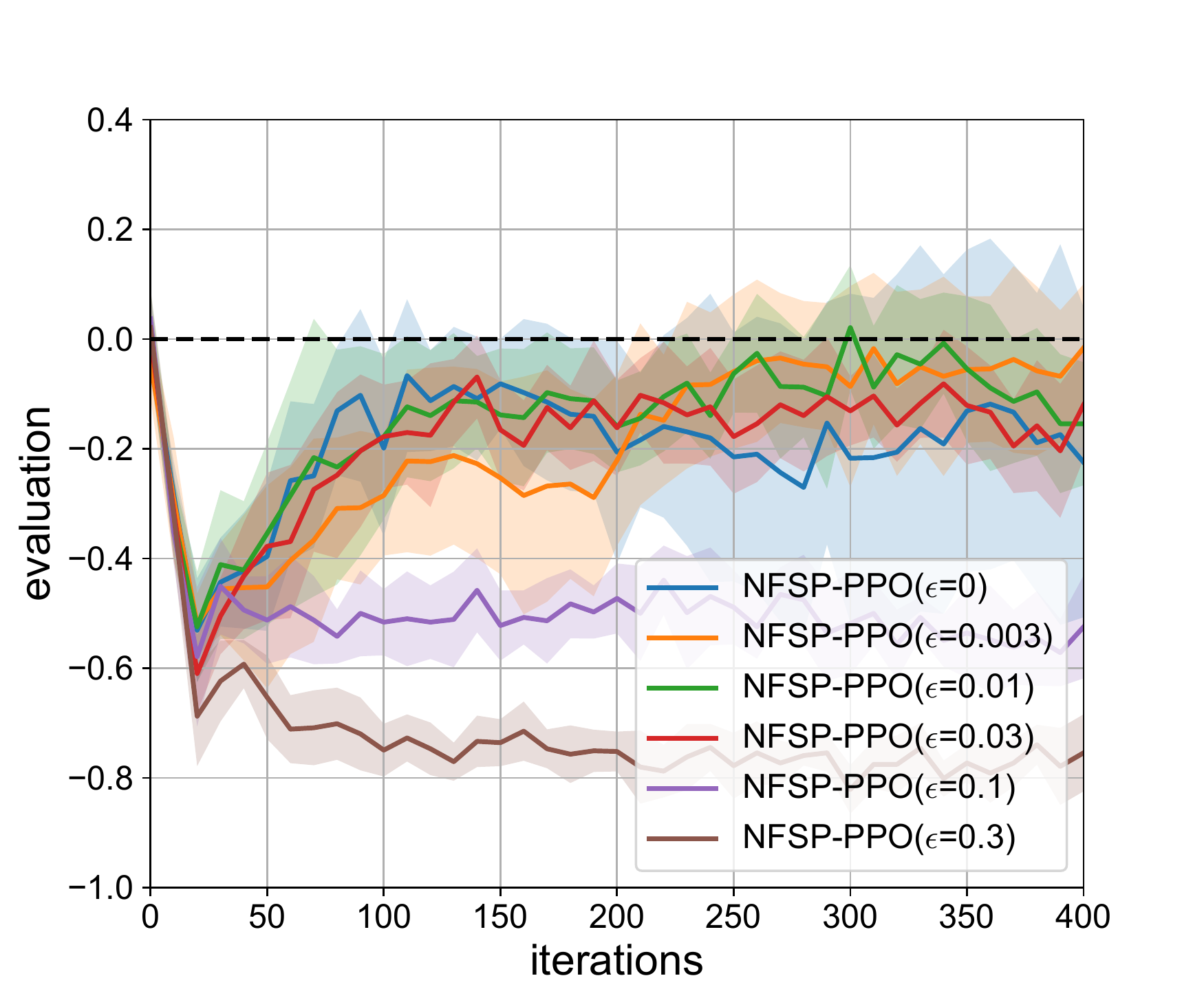}
  \caption{NFSP-PPO vs EPO}
  \end{subfigure}
  \begin{subfigure}[b]{.22\textwidth}
  \centering
  \includegraphics[width=1\textwidth, height=0.7\textwidth]{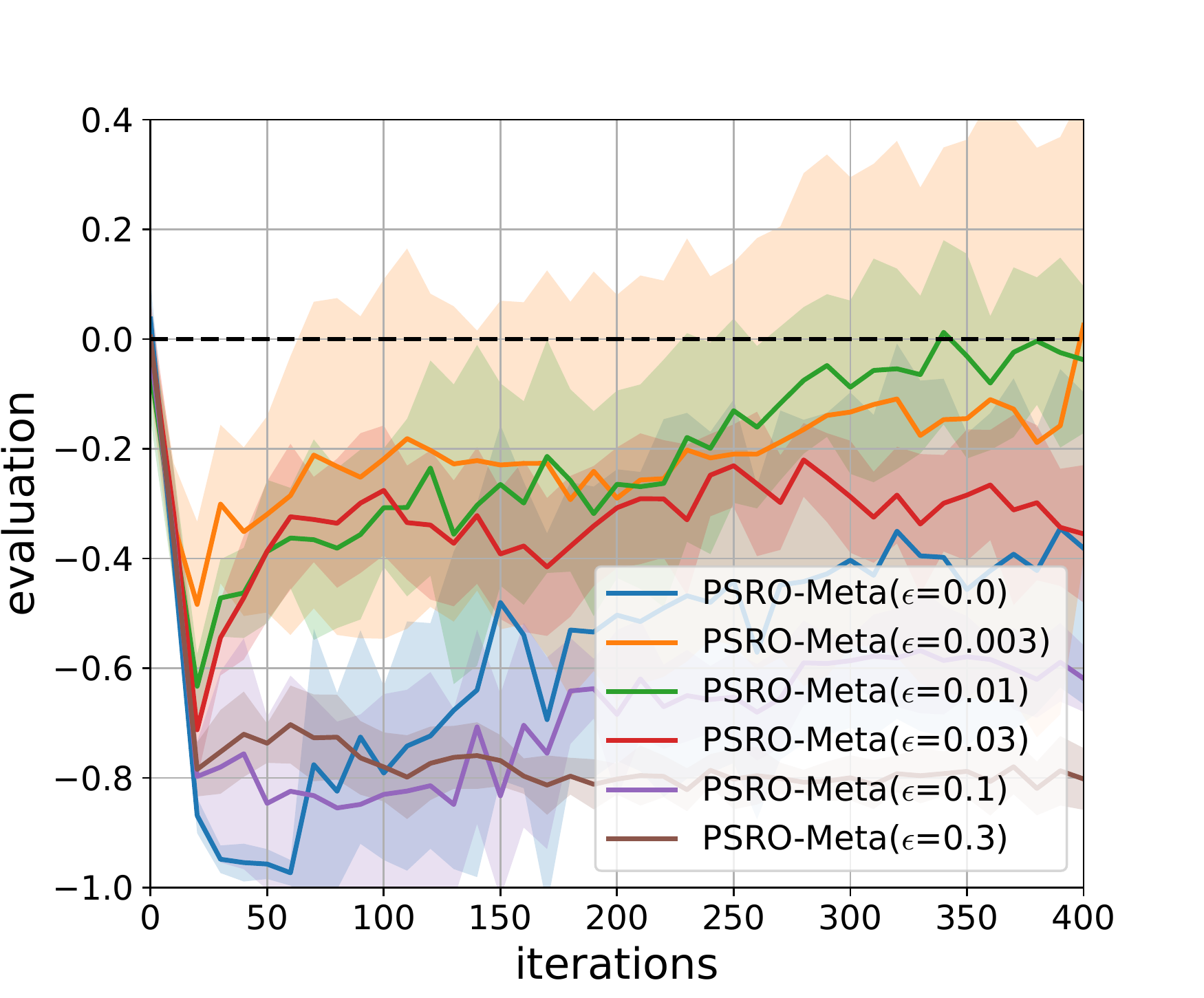}
  \caption{PSRO-Meta vs EPO}
  \end{subfigure}
  \begin{subfigure}[b]{.22\textwidth}
  \centering
  \includegraphics[width=1\textwidth, height=0.7\textwidth]{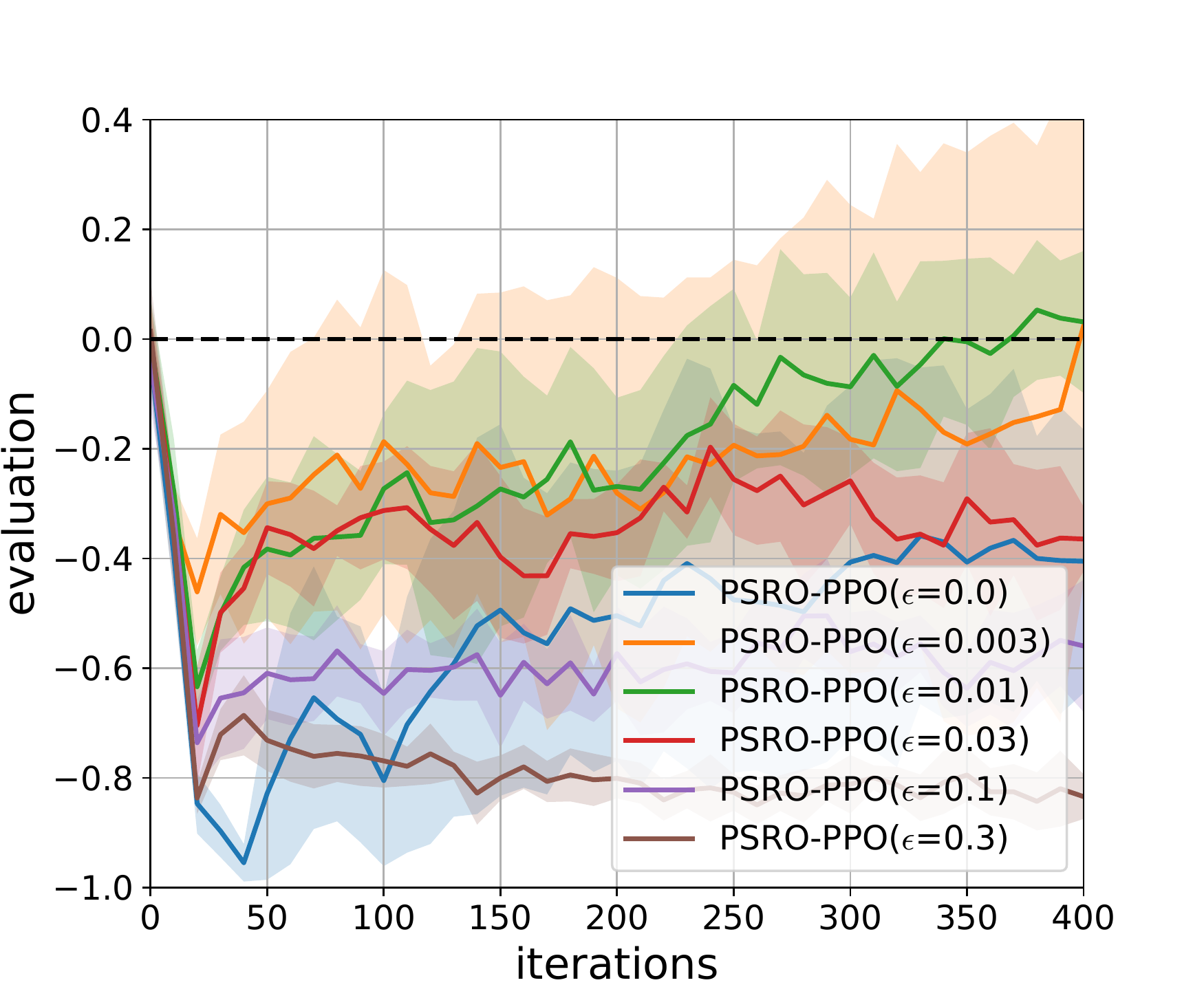}
  \caption{PSRO-PPO vs EPO}
  \end{subfigure}
  \caption{Performance of SP, NFSP, PSRO, PPO, and SimpleAI relative to baseline EPO($\epsilon=0.1$) along learning process. The curves are averaged over random seeds with solid lines indicating mean values and shadow areas indicating standard variance.
  NFSP-Fic indicates the fictitious player and NFSP-PPO indicates the fellow PPO agent.
  PSRO-Meta indicates the meta-strategy and PSRO-PPO indicates the fellow PPO agent.
  }
  \label{fig_diff_agents}
\end{figure}

The curves of relative performance are plotted in Figure \ref{fig_diff_agents}, and a common phenomenon is that all curves immediately drop below zero once the learning starts. It indicates no algorithm improves policies as fast as EPO, and in other words, EPO is advantageous in finding policy gradient towards Nash equilibrium.
If only playing against a fixed opponent, PPO agents are not possible to approach Nash equilibrium, reflecting low relative performance against EPO. With the increase of iterations, SP, NFSP-PPO, and PSRO-PPO agents with specific entropic parameters are able to close the gap to EPO. One probable reason is that Wimblepong game does not severely suffer from strategic cycles \citep{david2019open}, so even for simple SP,
it is possible to approach Nash equilibrium by beating ever-improving opponents.
The fictitious player of NFSP makes much slower progress than its PPO fellow. It demonstrates that supervised learning is less efficient in long-term decision-makings than reinforcement learning.

\begin{figure}
  \centering
  \begin{subfigure}[t]{.1063\textwidth}
  \centering
  \includegraphics[width=1\textwidth, right]{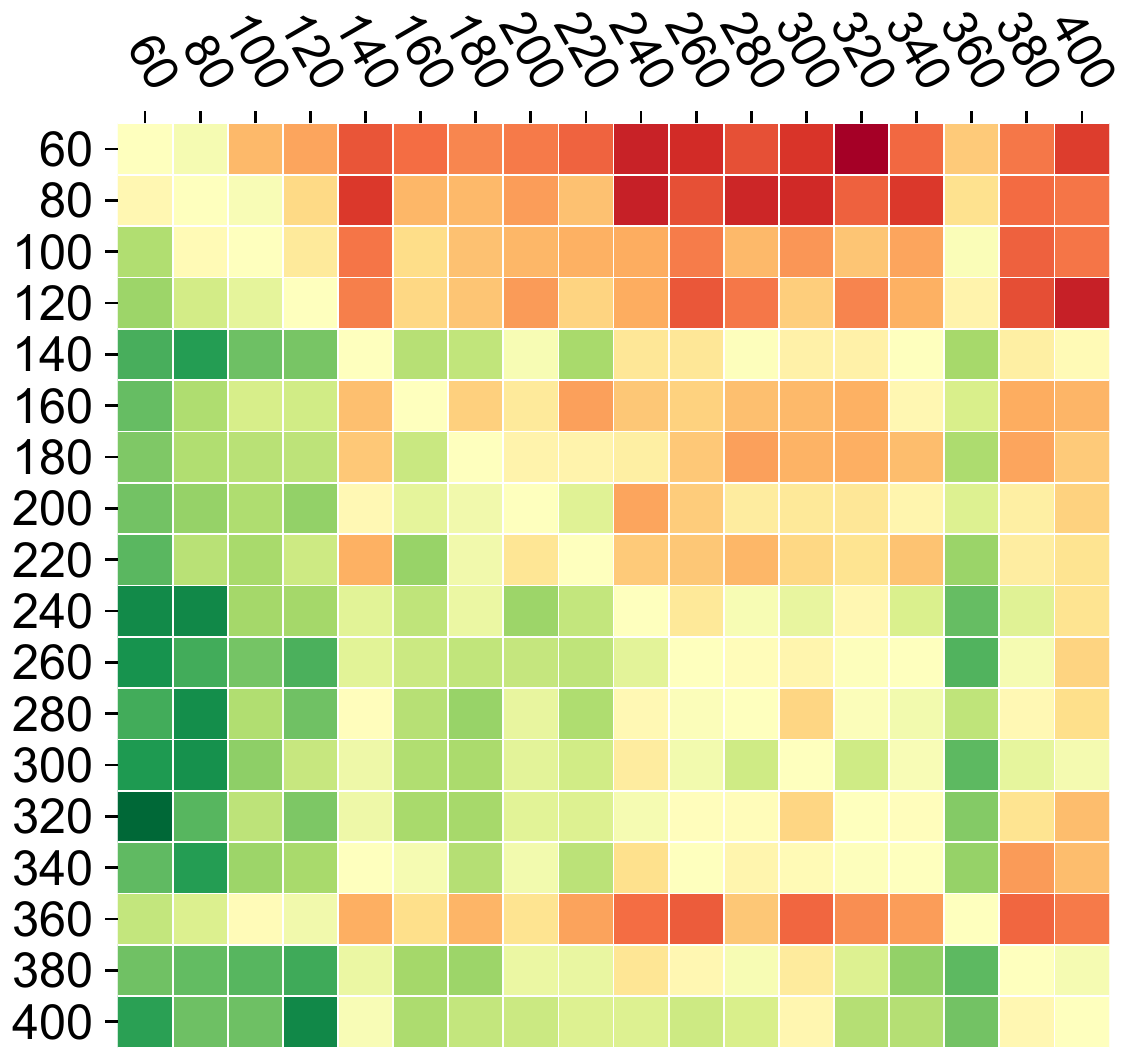}
  \end{subfigure}
  \begin{subfigure}[t]{.1063\textwidth}
  \centering
  \includegraphics[width=1\textwidth, right]{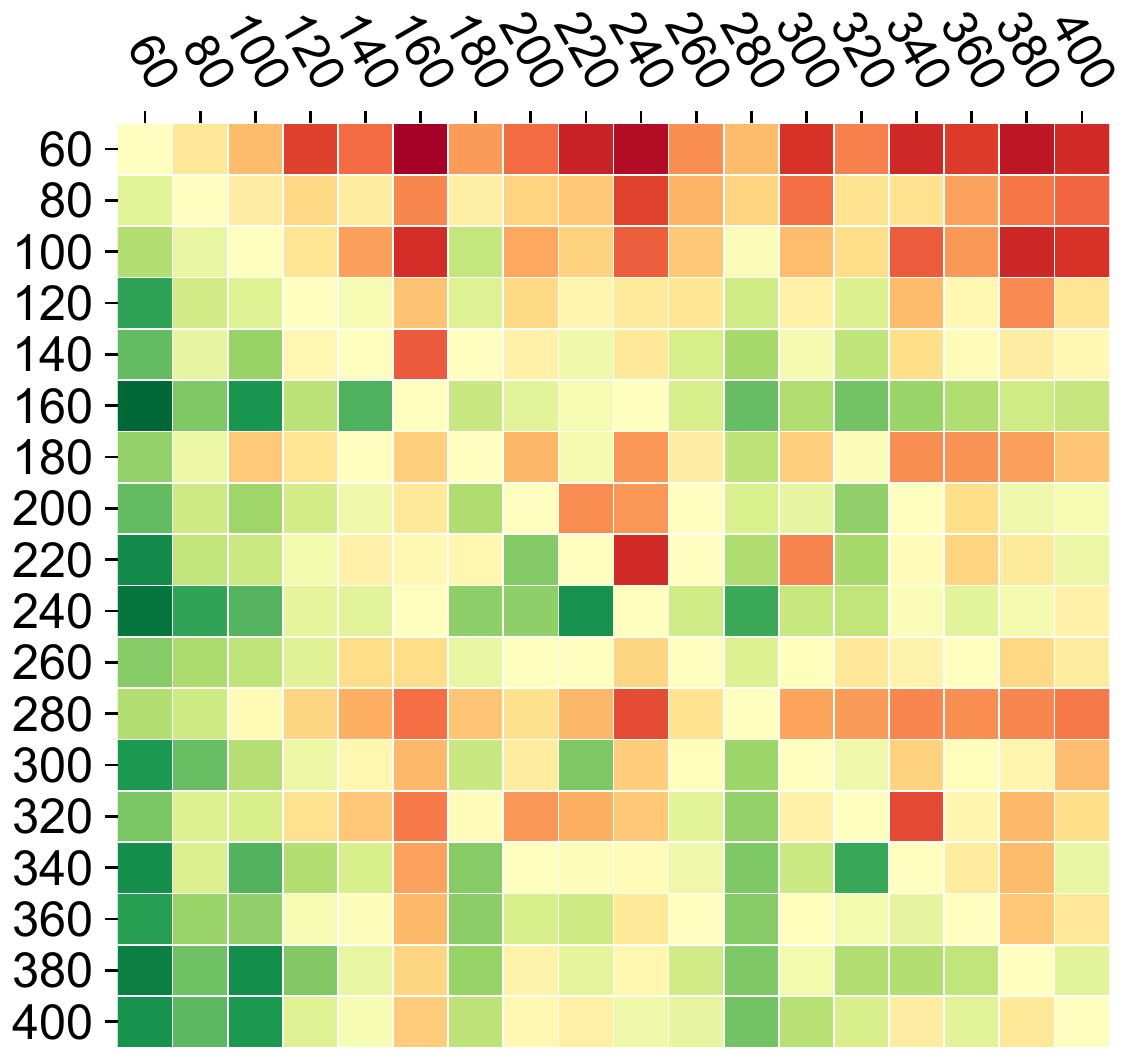}
  \end{subfigure}
  \begin{subfigure}[t]{.1063\textwidth}
  \centering
  \includegraphics[width=1\textwidth, right]{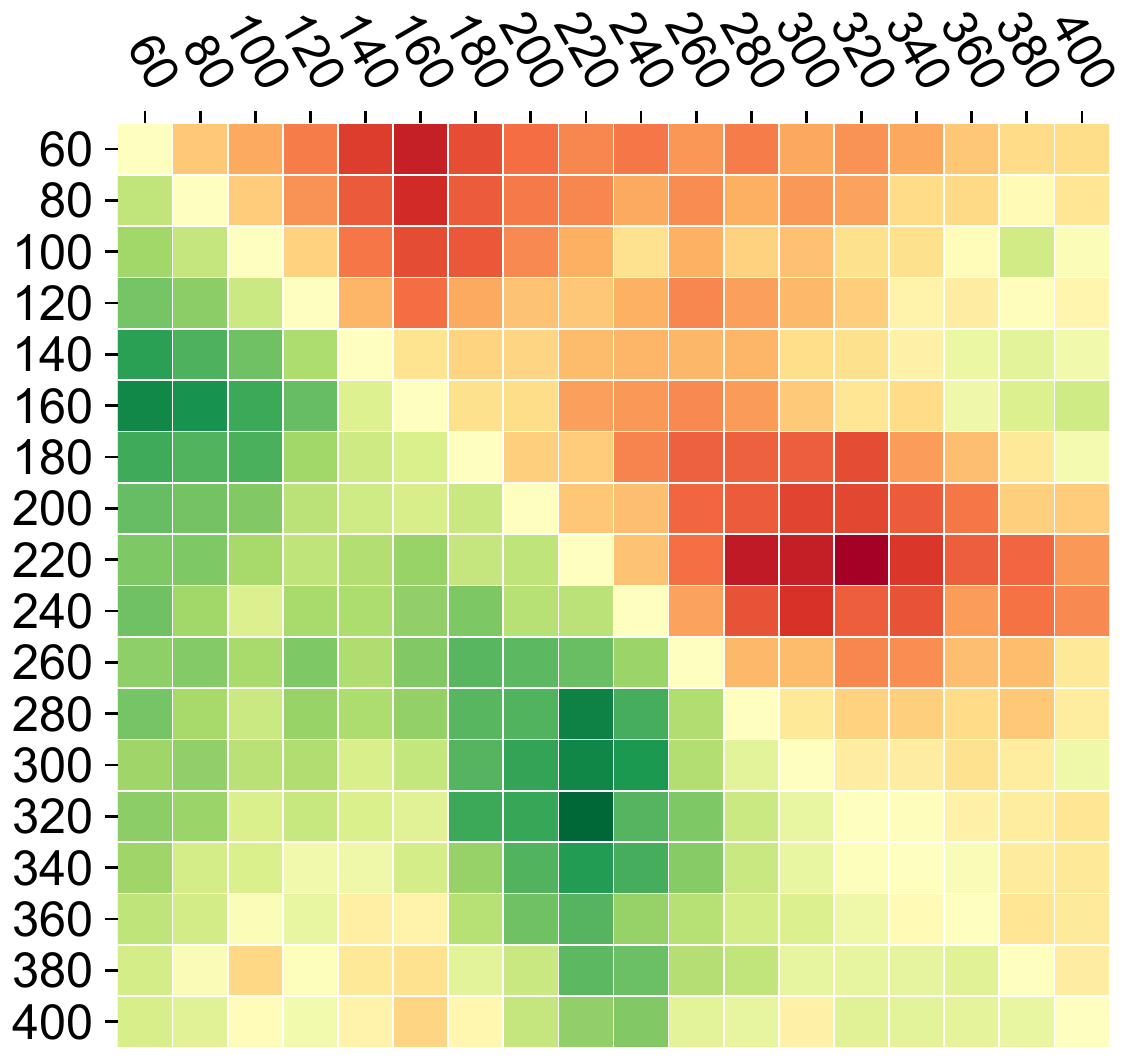}
  \end{subfigure}
  \begin{subfigure}[t]{.127\textwidth}
  \centering
  \includegraphics[width=1\textwidth, right]{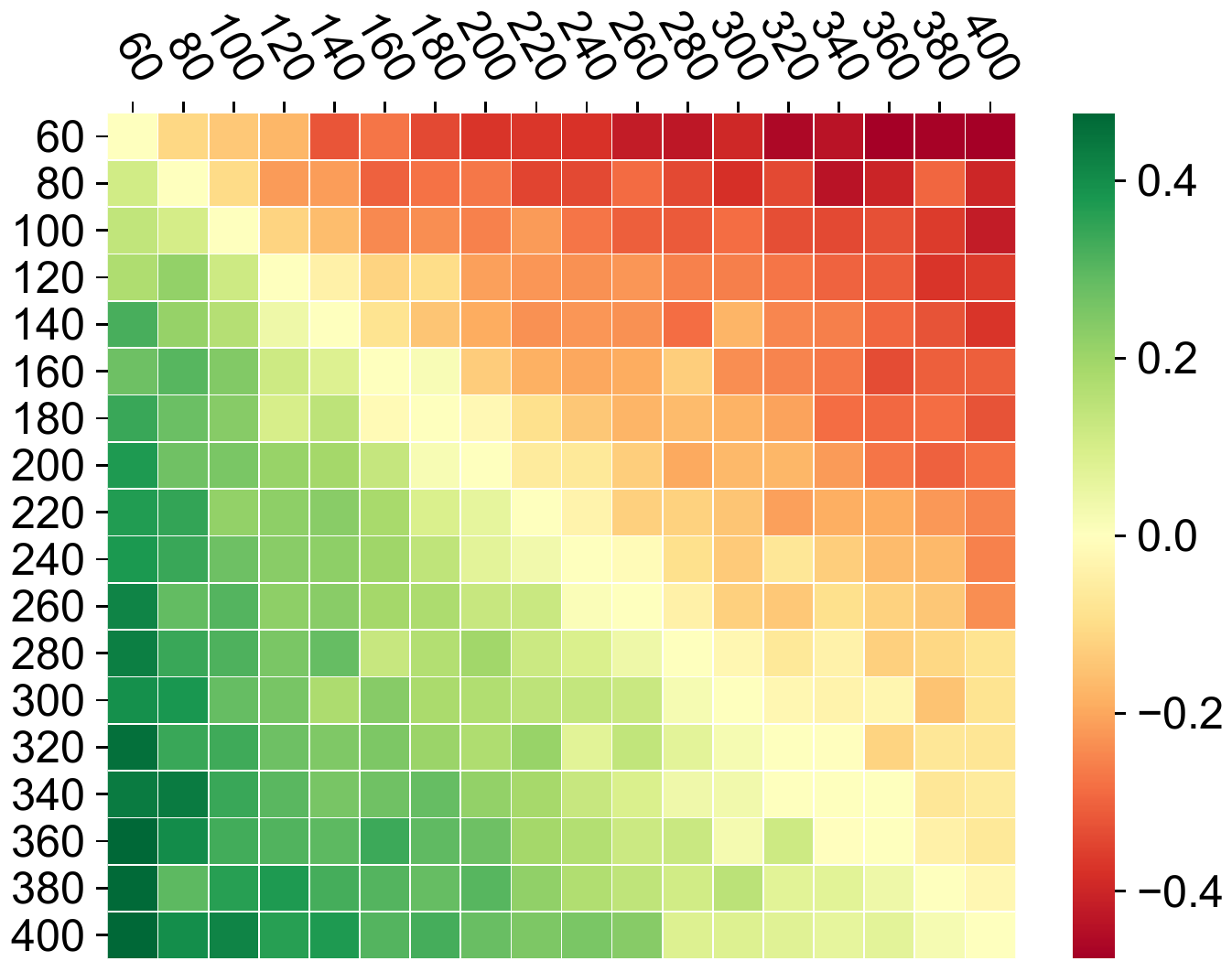}
  \end{subfigure}
  \\ \vspace{0em}
  \begin{subfigure}[t]{.1063\textwidth}
  \centering
  \includegraphics[width=0.9\textwidth, left]{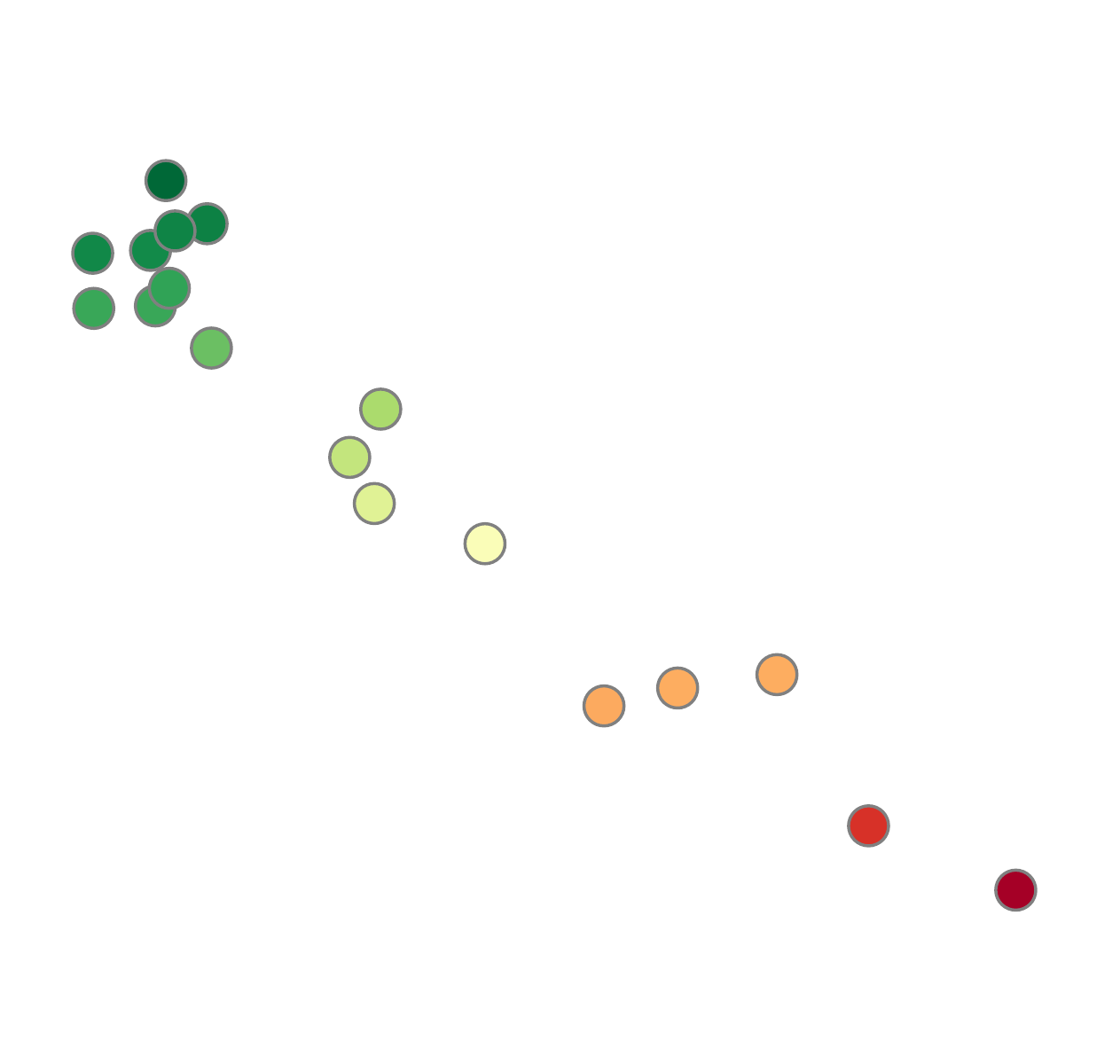}
  \caption{ }
  \end{subfigure}
  \begin{subfigure}[t]{.1063\textwidth}
  \centering
  \includegraphics[width=0.9\textwidth, left]{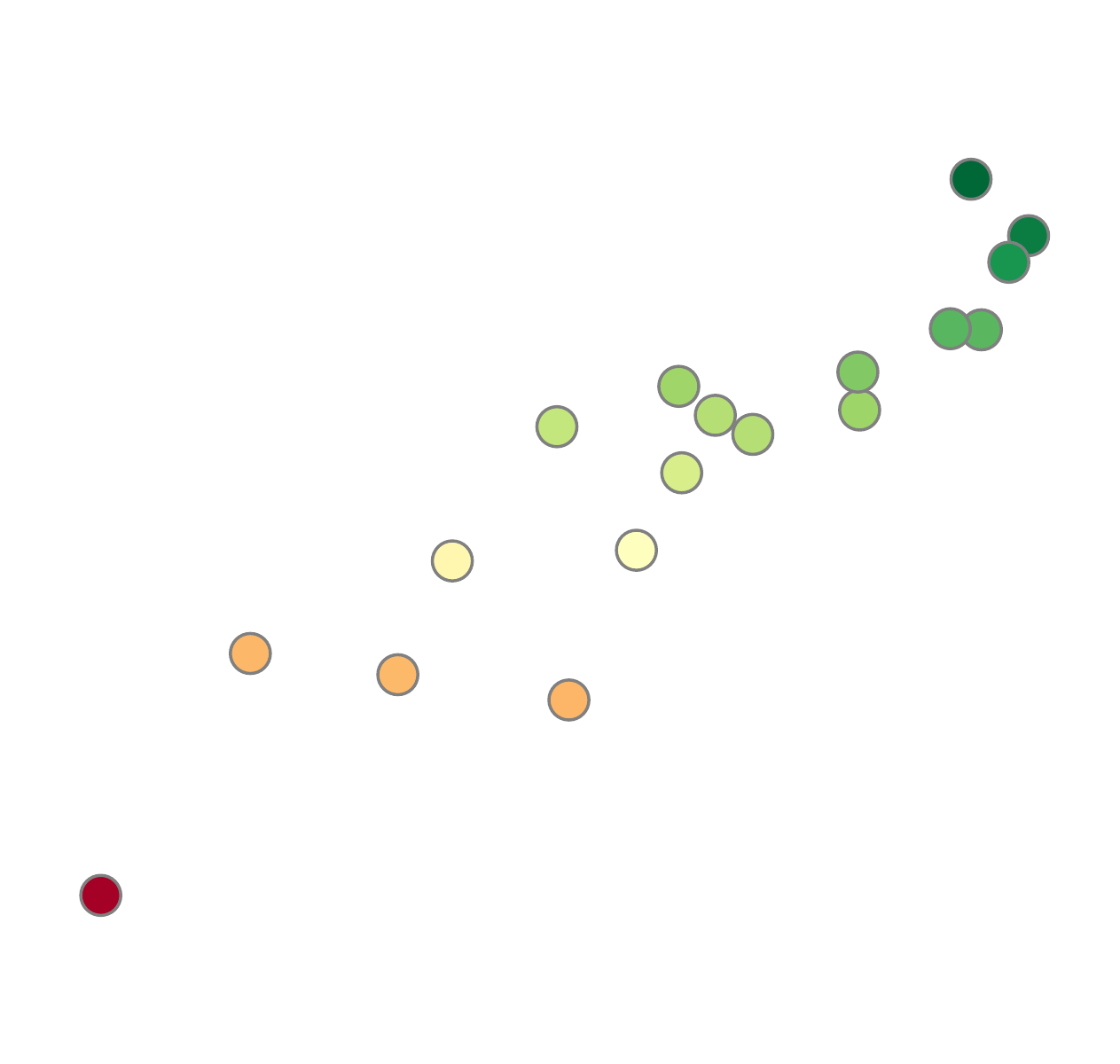}
  \caption{ }
  \end{subfigure}
  \begin{subfigure}[t]{.1063\textwidth}
  \centering
  \includegraphics[width=0.9\textwidth, left]{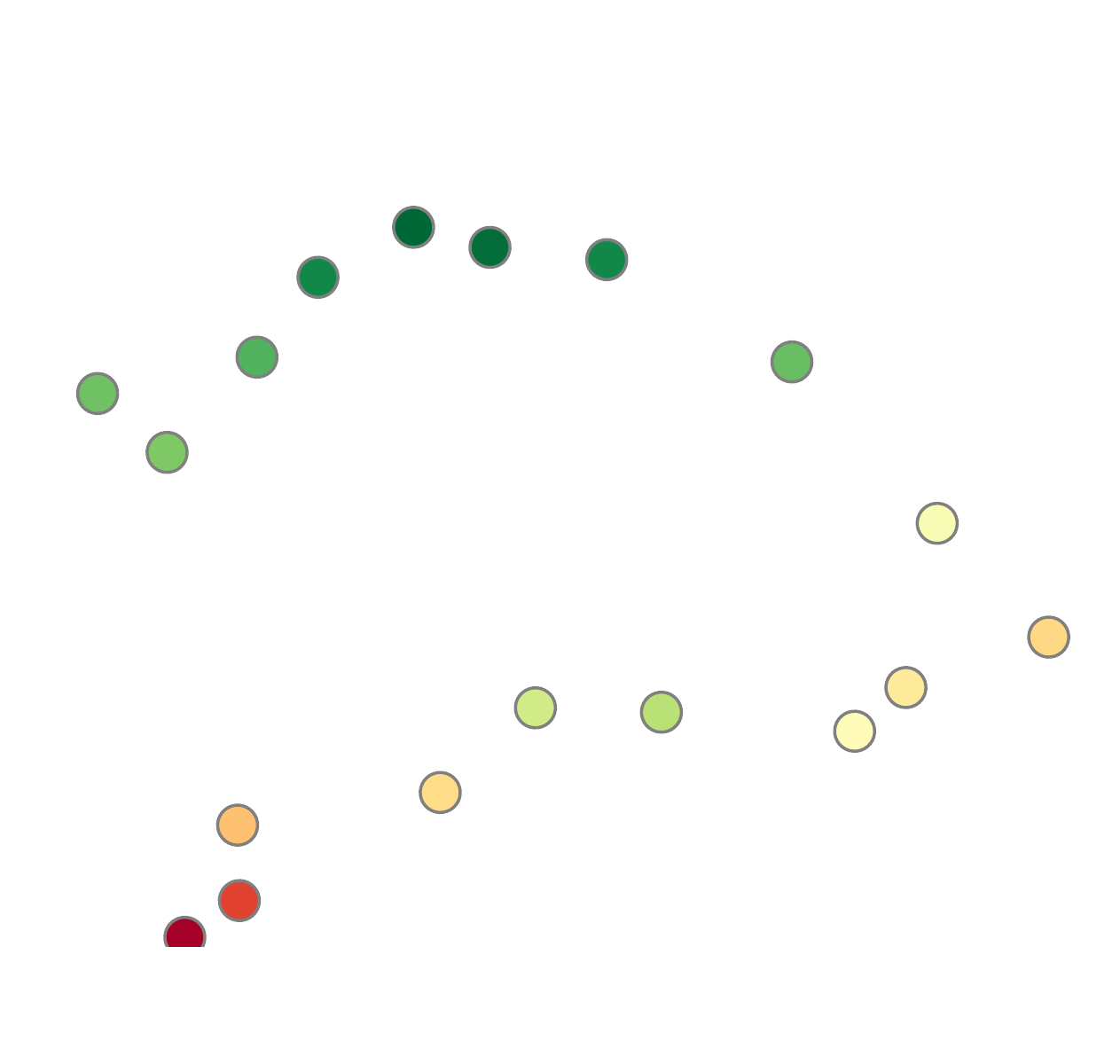}
  \caption{ }
  \end{subfigure}
  \begin{subfigure}[t]{.1063\textwidth}
  \centering
  \includegraphics[width=0.9\textwidth, left]{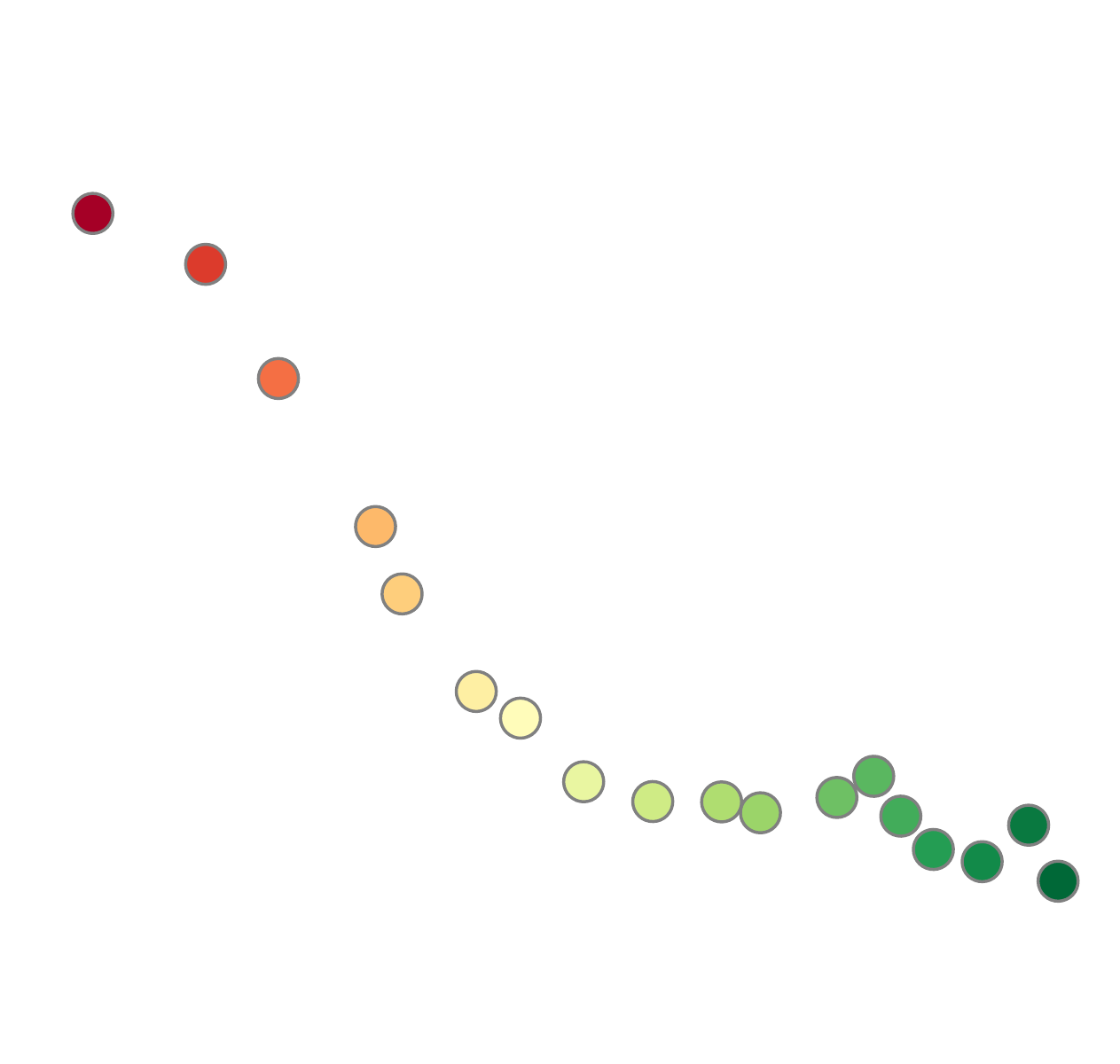}
  \caption{ }
  \end{subfigure}
  \caption{Payoff tables and visualization of EPO populations under different experience buffers. (a): Latest 1-iteration replay, (b): Latest 10-iteration replay, (c): Latest 100-iteration replay, (d): Full historical replay. \textbf{Top}: Row and column labels indicate agents at different iterations. \textbf{Bottom}: 2D embedding of payoff table by using the first 2 dimensions of Schur decomposition; Color corresponds to average payoff of an agent against entire population. }
  \label{fig_wimble_epo_in_replay}
\end{figure}

\paragraph{Ablation study} We now investigate the effect of historical experience in EPO and run experiments with different sizes of experience buffer. Note that when the buffer stores only experience of the latest iteration, the algorithm becomes self-play. Agents after different iterations in an experiment forms a population and their payoff table is drawn in Figure \ref{fig_wimble_epo_in_replay}. We also plot the 2-dimensional visualization by Schur decomposition \citep{david2019open} at the bottom of the figure. Full replay of historical experience makes EPO update policies in a transitive or monotone mode. Limited replay makes algorithm suffer from policy forgetting, in the sense that new policies may forget how to beat some old policies in history. It corresponds to cyclic or mixed shapes in the 2D embedding of policy populations.

\section{Conclusion}

A game-theoretic learning framework for $n$-player Markov games is proposed in this paper. The convergence of the dynamical learning system to an approximate Nash equilibrium is proved by Lyapunov stability theory, and is also verified on different $n$-player MG examples. The combination of NNs makes the EPO algorithm applicable to large games. The distributed implementation and no need of game interactions with specific opponents makes it appealing to companies and groups that are less intensive in computing resources.

There is still space for improvement. Existence of multiple Nash equilibria may pose a risk to our work, leading to the decrease of social welfare. Correlated equilibrium \citep{farina2020co, celli2020no} is a potential solution. We encourage research to investigate how small a common knowledge can be introduced to achieve a promising outcome in coordination games.

%There is still space for improvement, especially considering the simplicity and practical success of SP in board games and other tasks. One direction is to prioritize historical experience \citep{2016Prioritized} and evolve a group of players for diversity \citep{david2019open, Nieves2021ModellingBD}.
%Existence of multiple Nash equilibria may pose a risk to our work, leading to the decrease of social welfare. Correlated equilibrium is a potential solution. We encourage research to investigate how small a common knowledge can be introduced to achieve a promising outcome in coordination games. In addition, our training relies on the entire historical experience and hence makes a requirement of memory space, which may cause the increase of hardware costs in large games.

\clearpage
\bibliography{ref}

\end{document}